\documentclass{article}
\usepackage{amssymb}
\usepackage{amsmath,amsthm}
\usepackage{amsfonts}
\newcommand{\cc}{\mathbb{C}}

\newcommand{\ip}{{\,{\rm i}\,}}

\newcommand{\nn}{\nonumber}

\newcommand{\bee}[1]{\renewcommand{\baselinestretch}{1}\begin{equation}\label{#1}}
\newcommand{\ee}{\end{equation}\renewcommand{\baselinestretch}{2}}
\newcommand{\ba}{\left[\begin{array}}
	\newcommand{\ea}{\end{array}\right]}
\newcommand{\bac}{\begin{array}}
	\newcommand{\eac}{\end{array}}
\newcommand{\eacb}{\end{array}\!\!\!\!\!\!}
\newcommand{\bc}{\begin{center}}
\newcommand{\ec}{\end{center}}
\newcommand{\beaa}{\renewcommand{\baselinestretch}{1}\begin{eqnarray}}
\newcommand{\eeaa}{\end{eqnarray}\renewcommand{\baselinestretch}{2}}

\newcommand{\half}{\mbox{$\frac{1}{2}$}}
\begin{document}
\begin{titlepage}
\thispagestyle{empty}
\title{{\bf Lorentz--like extension Mixing Higgs and Gauge Fields in a U(1) symmetric model}}

\author{Luis Alberto {\bf Wills-Toro}$^{1,2,3}$\\
https://orcid.org/0000-0001-8815-8782\\
~\\
1. School
 of Mathematics, Universidad Nacional de Colombia,\\ Medell\'{\i}n,
 Colombia, e--mail: lawillst$@$unal.edu.co\\
2. Departamento de F\'{\i}sica Te\'orica y del Cosmos, Universidad de Granada\\
3. CIMAT}
%\date{}
\maketitle
\begin{abstract}
  A program  searching for symmetry structures behind  some features of the standard Model is launched. After addressing known no-go theorems, we construct a novel symmetry mixing gauge and Higgs fields which is a Lorentz symmetry extension involving  gauge symmetries. We construct a  field theory model with such a local symmetry, which is in turn invariant under local Lorentz transformations.  We address the  action of extended symmetries on the geometric frame, and models with extra dimensions. This features the novelty of diverse covariant derivatives associated to non--commuting differential operators, and their associated curvatures. Mass--like terms arise accompanying non--commuting differential operators. A novel formalism for constructing an invariant Lagrangian in direct correspondence to obtained casimir operators --embodying a long sought paradigm-- is developed. It recovers  both gauge as well as some sort of Higgs-like potential. Equations of motion and conserved currents are found. Some exploration is devoted to possible invariant measures hinting possible approaches to address anomalies.

~
{\bf  }
\end{abstract}
\end{titlepage}
\pagestyle{myheadings} \markright{\hfill L. A. Wills-Toro: Mixing Higgs and Gauge Fields...\ \ \ \ \ \ \ }

 \pagenumbering{arabic}
\section{Introduction}
We aim to initiate a program focused on the exploration of novel symmetries within Quantum Field Theory (QFT) models. The objective is to shed light on various challenges, including but not limited to dark energy, dark matter, neutrino masses, scale naturality, and background independence, thereby extending beyond the confines of the Standard Model (SM) of particle physics.

Outlined below are the proposed guiding principles for this endeavor:

\begin{itemize}
	\item[$\bullet$]
 We aspire to introduce a sense of naturality to the somewhat ad hoc structures of the SM, surpassing the conventional "internal $\oplus$ external" symmetry paradigm.
	\item[$\bullet$]
 An emphasis is placed on streamlining particle standard brands —Gauge, Matter, or Higgs. It is desirable that all particles and fields, whether bosonic or fermionic, correlate with specific symmetry generators or their remnants in the low-energy domain. We advocate for Lagrangian densities derived from invariants of the considered symmetries.
	\item[$\bullet$] 
 Recognizing established physical and mathematical structures as potentially emerging or effective outcomes of yet to uncover structures of earlier stages of the universe, we seek to investigate the nature of these structures. This encompasses the consideration of minimal couplings, connections, Wick rotations, causality, spontaneous symmetry breaking, inertial effects, and more as emergent or effective  phenomena.
	\item[$\bullet$] 
 The constraints imposed by well-known No-Go theorems should be meticulously understood. These constraints open avenues for considering symmetries that may spontaneously break and no longer persist as S--Matrix symmetries in subsequent stages.
	\item[$\bullet$]
 Exploring potential connections between space--time and internal symmetries is a key objective. Initially, we might treat space--time coordinates just as parametrizations of the 4D manifold, with no evident relationship to the symmetries introduced through connections.	
	\item[$\bullet$]
 We anticipate symmetries that remain invariant under space-time-depen\-dent transformations. There is no initial assurance that these symmetries can be non--trivially represented  through plain differential operators in space-time coordinates, we remain open to suggestions of relations to theories in greater superspaces or with extra dimensions. Considering the challenges posed by higher spin fields, simple groups with non--positive definite metrics, and non-compact or non-simple groups, we acknowledge the need for a careful study under new settings, extending beyond well-known no-go theorems, to determine if they lead to consistent QFT models.
	\item[$\bullet$] 
 We anticipate that, through spontaneous symmetry breaking or other mathematical constructions, we may arrive at diverse structures within the standard model and its coupling to classical gravity. The descent from more restrictive symmetries may reveal emerging space-time symmetries through differential operators in space-time, enriching both the geometry of the manifold through fields and the dynamics of the fields through the enriched geometry. Certain symmetries may persist only as global symmetry remnants.
	\item[$\bullet$] 
 An objective is to ponder the reunion of hitherto unrelated particles into multiplets of novel symmetries, thereby combining internal and external symmetries. Given that all particles in the SM fall into fundamental (for matter and Higgs) or adjoint (for gauge fields) representations of the Lorentz and internal symmetries, a consideration of multiplets involving gauge and Higgs, and super--multiplets involving matter with gauge or Higgs fields is in order.
\end{itemize}
It would be desirable to implement a mechanism ensuring that, in addition to the Electro-Weak (EW) gauge bosons, the EW scalars remain massless upon reaching energy scales that initiate EW symmetry breaking \cite{GW}. This mechanism should consistently provide masses that subsequently decouple additional gauge fields in the low-energy domain. Both the Higgs and the electroweak gauge fields in the Standard Model (SM) exhibit analogous ultraviolet features and become entangled in a nonlinear recombination of degrees of freedom following spontaneous symmetry breaking.

We consider it plausible that at higher energies, a certain symmetry may have governed a mixing among the Higgs field $H_{\alpha}$, $\bar{H}^{\alpha}$, the isospin gauge field  $A_{[i]\mu}$ (for $SU(2)$ with hypercharge 0, where $i$ is the adjoint $SU(2)$ representation index and $\mu$ is the space-time index, with $g_{\mu\nu} = {\rm diag}(1,-1,-1,-1) = g^{\mu\nu}$), and the hypercharge gauge field $B_{\mu}$ (for $U(1)$). As these fields have a naive energy dimension of 1, we introduce dimensionless  symmetry generators $T_{\alpha\mu}$ and $\hat{T}^{\,\,\alpha}_{\mu}$ combining gauge and Higgs fields:
\beaa
&&[T_{\alpha\mu},\bar{H}^{\beta}]\!=\!a'\sigma[i]_{\alpha}^{\,\,\beta} A_{[i]\mu}\!+\!b'\delta_{\alpha}^{\,\,\beta} B_{\mu}\!+\!\cdots,\ [T_{\alpha\mu},B_{\nu}]\!=\!c'g_{\mu\nu}{H}_{\alpha}+\cdots,\label{THbar}\\ 
&&[\hat{T}_{\mu}^{\ \alpha}\!, {H}_{\beta}]\!=\!-\bar{a}'\sigma[i]_{\beta}^{\,\,\alpha} A_{[i]\mu}\!-\!\bar{b}'\delta^{\ \alpha}_{\beta} B_{\mu}\!+\!\cdots,\  [\hat{T}^{\,\,\alpha}_{\mu},B_{\nu}]=-\bar{c'}g_{\mu\nu} \bar{H}^{\alpha}+\cdots,\ \ \ \label{TB}\\
&&[T_{\alpha\mu},A_{[i]\nu}]\!=\!f'g_{\mu\nu}\sigma[i]_{\alpha}^{\,\,\beta} {H}_{\beta}\!+\!\cdots,\ [\hat{T}^{\,\,\alpha}_{\mu},A_{[i]\nu}]=-\bar{f}'g_{\mu\nu} {\bar{H}}^{\beta}\sigma[i]_{\beta}^{\ \ \alpha}\!+\!\cdots.\ \ \ \label{TA}
\eeaa
where $a',b', c',f'$ are constants. Phenomenologically, there are indications of a nontrivial relationship between internal and external symmetries. For example, leptons, as particles, exhibit  spin multiplets (external symmetry) components that respond differently under electroweak interactions. Employing representation theory, we investigate the closure of the naive energy dimension 0 subalgebra, which includes generators $T_{\alpha\mu}$ and $\hat{T}^{\ \alpha}_{\mu}$. This subalgebra may encompass the Lorentz subalgebra with generators $M_{\mu\nu}$ or a similar entity, a real Lorentz symmetric traceless tensor  $T_{\mu\nu}$ (whose role is yet to be specified), a scalar generator $D$, as well as  generators $T_{[i]}$ of isospin and ${T_{[o]}}$ of hypercharge:
\beaa
\,[T_{\alpha\mu},\hat{T}^{\,\,\beta}_{\nu}]&=&-\ip a\delta_{\alpha}^{\,\,\beta}M_{\mu\nu}+u\delta_{\alpha}^{\,\,\beta}T_{\mu\nu}\nn\\& &-q g_{\mu\nu}\delta_{\alpha}^{\,\,\beta} D- b  g_{\mu\nu} \sigma[i]_{\alpha}^{\,\,\beta}T_{[i]}+ (h/2) g_{\mu\nu}\delta_{\alpha}^{\,\,\beta}{T_{[o]}},\label{TTbar}
\eeaa
where $a,u,q,b,h$ are constants. 
This appears to resemble an extension of conformal symmetry. Relation (\ref{TTbar}) is the Ansatz  that such a commutator can be expressed as a linear combination of an antisymmetric Lorentz tensor $M_{\mu\nu}$, a traceless--symmetric Lorentz tensor $T_{\mu\nu}$, an isospin triplet $T_{[i]}$, and so forth. Extensions of this kind, involving both internal and external symmetries, have been comprehensively explored under constraints imposed by:\\

{\bf{Theorem}} [Coleman--Mandula No-Go]\cite{ColMan} {\it
If $G$ is a symmetry group for the $S$-Matrix of a QFT model, where:
{\bf (i)} For any $M>0$, there are only a finite number of particle types with mass less than $M$;
{\bf (ii)}  Scattering occurs at almost all energies (except for perhaps some isolated set of energies);
{\bf (iii)} The amplitudes for elastic two-body scattering are analytic functions of the scattering angle at almost all energies and angles;
then the generators of $G$ consist only of the generators of the Poincaré group $P$ and scalar  internal symmetry generators.}\\

However, there are instances where certain constraints imposed by the Cole\-man-Mandula (CM) theorem can be circumvented \cite{SL}. Notably, spontaneously broken symmetries within the action, once broken, are not obliged to act on the $S$-Matrix or necessarily commute with it. Additionally, discrete symmetries remain admissible.

In the context of massless theories, certain arguments concerning antisymmetry in the CM proof may not hold, as seen in the case of conformal symmetry \cite{MZ}\cite{AlbaDiab}. The CM proof relies on the use of commutation relations, and a broader algebraic framework can potentially evade the theorem. This is evident in the case of $\mathbb{Z}_2$
--graded or Lie superalgebras, as elucidated by the Haag--{\L}opusza\'nski--Sohnius no-go theorem \cite{HLS}. Moreover, the condition of analyticity of $S$ with respect to the scattering angle of two bodies in 1+1 dimensions is not fulfilled, given the presence of only forward and backward scattering.

Our investigation encompasses an initial stage involving solely massless fields and a symmetry that enlarges the Lorentz algebra, subsequently extending to not-necessarily massless fields due to non--commutive covariant derivatives in both the enlarged Lorentz and Poincaré algebras. We anticipate that spontaneous symmetry breaking or some other emerging mechanism may lead to a model aligning with the constraints imposed by the CM theorem or its corresponding generalization. Our approach involves extracting insights from the obtained outcomes to guide the extension of familiar structures.

The key methodologies we endorse include the Lie algebraic closure of the proposed structures, exploration of their invariants, consideration of hypothetical covariant derivatives, examination of associated curvature, and the construction of invariant Lagrangian densities linked to Lie algebraic invariants. We systematically label and group sets of generators associated with specific (irreducible) representations under the subgroups prominent in the low-energy realm (Lorentz and internal symmetries). The  separate labeling for their structure constants facilitates the tracing of diverse contributions to the model.

Given that the actual nature of the Lie algebraic structures is initially unknown, questions regarding simplicity or semisimplicity, as well as the quest for independent couplings for their simple factors, remain unresolved. To foster and encourage the innate ingenuity of our approach, we postpone the exploration of relations and overlaps with various well-established advancements in this and related domains until the final stages of our inquiry.

A crucial point for discussion remains—the Higgs field $H_{\alpha}, \bar{H}^{\alpha}$
 resides in the fundamental representation of the isospin symmetry. Consequently, to accommodate a Higgs field as a gauge field, we must permit a symmetry that includes generators in the fundamental representations of $SU(2)$. The reader may discern a parallelism with the pursuit of supersymmetry, involving the fundamental representations of the Lorentz symmetry within the extended symmetry. Subsequent stages of this program will address this particular undertaking.

The transformations of fields in (\ref{THbar})--(\ref{TTbar}) encompass both space-time and internal symmetry groups, potentially leading to a nontrivial connection between internal and external symmetries. Here, we develop this program first in the absence of the $SU(2)$ symmetry to explore a simpler case, thus eliminating the indices $\alpha, \beta, i, \ldots$. A forthcoming contribution will address the case with $SU(2)$, involving a fundamental representation thereof.

Our exploration takes us along a path to discover truly novel extensions of the Lorentz algebra. In Section 2, we determine a Lie algebra that allows for a mixing between internal gauge connections and Higgs fields. Moving to Section 3, we develop a formalism for a covariant derivative, aiming to obtain symmetry transformations not only for the connections and their curvature but also a formalism to derive an invariant Lagrangian. Section 4 delves into the determination of differential representations of the obtained extension of the Lorentz algebra, facilitating the definitive classification of the extension and recovering the previously identified invariants. This section represents the most direct route to classifying the involved Lie algebra.

Transitioning to Section 5, we revisit covariant derivatives, now involving the obtained differential representations, developing connections in a truly non-commutative setting as differential operators generally do not commute. Apart from obtaining symmetry transformations for such connections, this section determines corresponding curvatures and invariant Lagrangians. A notable result of this study is the emergence of curvature terms linear in the fields, leading to quadratic (mass--) terms in the invariant Lagrangian without violating symmetry. Additionally, a general formalism arises for obtaining invariant Lagrangians associated with the Casimirs of the corresponding Lie algebra (both for the Lorentz and internal symmetry indices linked to the Lie algebra generators, as well and separately for the Einstein indices associated with covariant derivatives). This addresses the quest for providing a less casuistic nature of the Lagrangians of connections of diverse spins, directly relating them to invariants, as posed in the Field Theory book by P. Ramond \cite{FTPremier}.

In Section 6, we complete the field theoretical construction by determining the equations of motion, the Gauge invariance of the Lagrangian, and identifying conserved currents. This section also explores the character of the theory as a field theory in an extended space-time, serving as the foundation for addressing the extended Poincaré in \cite{U1PoinExt}. Notably,  the field theory realizes connections on a Lie Manifold coordinatized with parameters that allow for the differential representation the symmetry.

A significant accomplishment of this research is the construction of a purely Gauge Model involving local external symmetries, such as Lorentz transformations. General relativity, in addition to implementing general diffeomorphisms invariance (through local translations), involves invariance under local Lorentz transformations through the spin connection. The second-order formalism expresses the latter in terms of the former. The relationship of the obtained  model to those of General Relativity will be addressed in the sequel of this study.

Section 7 addresses some alternative considerations, in particular, we give initial exploration of contributions to invariant measures, which seem to resonate with the study of anomalies. Section 8 provides concluding remarks.

\section{Extending the Dimensionless Subalgebras}
The closure of an algebra, including the Lorentz--like subalgebra for \(M_{\mu\nu}\) (in (\ref{LAlg})), the \(U(1)\) symmetry with generator \(T_{[o]}\), and the relation (\ref{TTbar}) dropping the \(SU(2)\) generator contributions, is based on assuming the following Ansätze: all generator indices are transformed as expected according to their Lorentz (as in (\ref{LTRinic}--\ref{LAlg})) and \(U(1)\) indices (as in (\ref{U1inic}--\ref{U1end})). We assign only non-trivial \(U(1)\) charge +1 to \(T_{\mu}\) and -1 to \(\hat{T}_{\nu}\). We do not establish yet if \(T_{\mu}\) and \(\hat{T}_{\nu}\) build an adjoint pair. We assume that Lorentz vectors are transformed into themselves under the action of \(T_{\mu\nu}\) (building an invariant subspace under \(T_{\mu\nu}\)). The coefficients (arrays) in the rhs-combinations are either invariants or associated with specific symmetry representations. After enforcing all Jacobi identities and suitable renormalization, the dimension 0 Lie algebra arises (writing only nontrivial relations):
 \beaa
 \,[T_{\mu},\hat{T}_{\nu}]&=&\epsilon(\ip M_{\nu\mu}+T_{\nu\mu}+(1/10) g_{\nu\mu} (4 {T_{[o]}}+{\hat{T}_{[o]}}))+2\epsilon{T_{[o]}},\label{Dim0Start}\\
  \,[{T_{[o]}},T_{\rho}]&=&T_{\rho},\ \ \ \ \ \ \ \ \ \ \  \,[{T_{[o]}},\bar{T}_{\rho}] =\ -\bar{T}_{\rho},\\
  \,[{\hat{T}_{[o]}},T_{\rho}]&=&T_{\rho},\ \ \ \ \ \ \ \ \ \ \  \,[{\hat{T}_{[o]}},\bar{T}_{\rho}]=\ -\bar{T}_{\rho},\label{U1inic}\\
  \,[T_{\mu\nu},T_{\rho}]&=&\sigma(\mu\nu)_{\rho}^{\ \tau}T_{\tau},\ \  \,[T_{\mu\nu},\bar{T}_{\rho}] =\ -\sigma(\mu\nu)_{\rho}^{\ \tau}\bar{T}_{\tau},\label{U1end}\\
   \,[M_{\mu\nu},T_{\rho}]&=&-\ip\sigma[\mu\nu]_{\rho}^{\ \tau}T_{\tau},\  [M_{\mu\nu},\bar{T}_{\rho}] =-\ip\sigma[\mu\nu]_{\rho}^{\ \tau}\bar{T}_{\tau},\label{LTRinic}\\
    \,[M_{\mu\nu},T_{\rho\sigma}]&=&-\ip(\sigma[\mu\nu]_{\rho}^{\,\,\tau}\delta_{\sigma}^{\,\,\omega}+\sigma[\mu\nu]_{\sigma}^{\,\,\omega}\delta_{\rho}^{\tau})T_{\tau\omega},\label{MinExStart}\\
      \,[M_{\mu\nu},M_{\rho\sigma}]&=&-\ip(\sigma[\mu\nu]_{\rho}^{\,\,\tau}\delta_{\sigma}^{\,\,\omega}+\sigma[\mu\nu]_{\sigma}^{\,\,\omega}\delta_{\rho}^{\tau})M_{\tau\omega},\label{LAlg}\\
   \,[T_{\mu\nu},T_{\rho\sigma}]&=&-\ip(\sigma(\mu\nu)_{\rho}^{\,\,\tau}\delta_{\sigma}^{\,\,\omega}-\sigma(\mu\nu)_{\sigma}^{\,\,\omega}\delta_{\rho}^{\tau})M_{\tau\omega},\ \  \rm{with}\\
   \sigma[\mu\nu]_{\rho}^{\tau}&:=&g_{\mu\rho}\delta_{\nu}^{\,\,\sigma}-g_{\nu\rho}\delta_{\mu}^{\,\,\sigma},\ \ 
\sigma(\mu\nu)_{\rho}^{\tau}:=g_{\mu\rho}\delta_{\nu}^{\,\,\sigma}+g_{\nu\rho}\delta_{\mu}^{\,\,\sigma}\!-\!\half g_{\mu\nu}\delta_{\rho}^{\,\,\sigma}\!,\ \ \label{MinExEnd}\\
 {\hat{T}_{[o]}}&:=&{T_{[o]}}+5{\cal T},\label{Dim0End} 
\eeaa
where \({\cal T}\) is some central charge of this extension. There is no short algorithm to determine the nature of a Lie algebra (beyond checking semisimplicity). We first explore its dimensions, subalgebras, and quadratic invariants to gain insight into its classification. Further below, we explore its differential representations, which provide a definitive classification of such a Lie algebra as a pseudo-unitary \({\mathfrak{u}}(s_1, t_1)\), where \(s_1, t_1 > 0\), and \(s_1 + t_1 = 5\).

The renormalization carried out can absorb some constants, but sometimes is unable to modify a sign. This is the origin of the parameter $\epsilon\in\{+1,-1\}$. Although we initially aimed for it to be a sign, we could eventually let $\epsilon\in\mathbb{R}$ as a free parameter that might decouple the obtained extension. We need to track distinctly positive and negative powers of $\epsilon$. We cannot interpret ${\hat{T}_{[o]}}$ as a dilatation, but $5{\cal T}={\hat{T}_{[o]}}-{T_{[o]}}$—which is a central charge of this subalgebra—could perhaps assume this role. The dimension 0 subalgebra (\ref{Dim0Start})--(\ref{Dim0End}) with generators $T_{\mu},\hat{T}_{\mu}, M_{\mu\nu},T_{\mu\nu}$, $(24 {T_{[o]}}+{\hat{T}_{[o]}})$ has dimension $24=5^2-1$ without its central charge ${\hat{T}_{[o]}}-{T_{[o]}}$ and $25=5^2$ with it. This suggests a possible relation respectively to the ${\mathfrak{su}}(t_1, s_1)$ and ${\mathfrak{u}}(t_1,s_1)$ with $s_1,t_1>0$ and $s_1+t_1=5$. The algebra of dimensionless generators (\ref{Dim0Start})--(\ref{Dim0End}) generated by $T_{\mu},\hat{T}_{\mu}, M_{\mu\nu},T_{\mu\nu}$, ${T_{[o]}}, {\hat{T}_{[o]}}$ will be called the Lorentz Extension (LE). The reason for maintaining both scalar generators ${T_{[o]}}, {\hat{T}_{[o]}}$ is that the central charge ${\hat{T}_{[o]}}-{T_{[o]}}$ of this Lorentz extension is no longer a central charge when considering a Poincaré-like extension. And the selection of these scalars as a basis will ensure that there exists a quadratic Casimir for both the Lorentz and Poincaré extensions that does not mix both scalar generators ${T_{[o]}}, {\hat{T}_{[o]}}$.

The subalgebra \(l = \langle \{M_{\mu\nu}, T_{\rho\sigma}\}\rangle\) in (\ref{MinExStart})--(\ref{MinExEnd}) has dimension \(15 = 4^2-1 = 6 \times \frac{5}{2}\), which hints at a possible relation to \({\mathfrak{su}}(t_2, s_2)\) with \(s_2, t_2 > 0\) and \(s_2 + t_2 = 4\), or to \({\mathfrak{so}}(t_3, s_3)\) with \(s_3, t_3 > 0\) and \(s_3 + t_3 = 6\) (the maximal Lie algebra of symmetries of a De Sitter \(t_3 = 1\) or Anti-de Sitter space \(t_3 = 2\)). Observe that there is a Klein geometry associated with this subalgebra, since the complement to the standard Lorentz algebra in this subalgebra \(l\) is invariant under the Lorentz algebra, building a reductive homogeneous space.

The subalgebra \(\langle \{M_{\mu\nu}, {\cal P}_{\rho}\}\rangle\), with \({\cal P}_{\rho}\) being \((T_{\rho}+\hat{T}_{\rho})/\sqrt{2}\) or \(\frac{\ip}{\sqrt{2}}(T_{\rho}-\hat{T}_{\rho})\) (notice \([{\cal P}_{\mu},{\cal P}_{\nu}] = \frac{\ip\epsilon}{\sqrt{2}} M_{\mu\nu}\)), has dimension 10 and is not isomorphic to the Poincaré algebra. It might be isomorphic to \({\mathfrak{so}}(t_o, s_o)\) with \(s_o, t_o > 0\) and \(s_o + t_o = 5\) (corresponding to a De Sitter \(t_o = 1\) for \(\epsilon = -1\) or Anti-de Sitter space \(t_o = 2\) for \(\epsilon = 1\)). There are actually two copies of the Poincaré algebra nested in the extension \(\langle \{M_{\mu\nu}, T_{\rho}\}\rangle\), and \(\langle \{M_{\mu\nu}, \hat{T}_{\rho}\}\rangle\). We do not confer to the generators \(T_{\rho}, \hat{T}_{\rho}\) the ultimate role of space-time translations, but they might contribute to the precursors of low-energy translations. In fact, we can consider a step towards contraction \(T_{\rho} \mapsto \Lambda {\cal T}_{\rho}, \hat{T}_{\rho} \mapsto \Lambda \hat{\bf \cal T}_{\rho}\), whose scale \(\Lambda\) might or might not be connected to the one that produces the contraction towards translations. We elaborate on this further under alternative considerations.
 
The  full  dimensionless algebra, LE, has  at least two casimirs, the central charge ${\hat{T}_{[o]}}-{T_{[o]}}$, and a  quadratic casimir $I$:
\beaa
I\!\!&:=&\!\!({2}/{\epsilon})\{\hat{T}^{\omega},T_{\omega}\}+M^{\rho\sigma}M_{\rho\sigma}+T^{\rho\sigma}T_{\rho\sigma}+(24/5){T_{[o]}}^2+(1/5)({\hat{T}_{[o]}})^2\nn\\
\!\!&&=[{T}^{\omega}\ \hat{T}^{\omega}\ M^{\rho\tau}\ T ^{\rho\tau}\ {T_{[o]}}\ {\hat{T}_{[o]}}]\cdot M\cdot [\hat{T}_{\omega}\ {T}_{\omega}\ M_{\rho\tau}\ T_{\rho\tau}\ {T_{[o]}}\ {\hat{T}_{[o]}}]^{tr}\!\!,\ \ \label{CasimirDim0}\\
&&{\rm with}\ \  M:={\rm diag}\left( \frac{2}{\epsilon}, \frac{2}{\epsilon}, 1, 
1, \frac{24}{5}, \frac{1}{5}\right)\ 
{\rm and}\  \{\hat{T}^{\omega},T_{\omega}\}:=\hat{T}^{\omega}T_{\omega}+T_{\omega}\hat{T}^{\omega}.\nn
\eeaa
In constructing this invariant, both the lowering and raising of Lorentz indices with \(g_{\mu\nu},\, g^{\mu\nu}\) and the matrix \(M\) (depending on the type of generator) come into play. Both contributions appear to construct a Killing Metric. Whether such a metric can be positive definite in a sense that allows for a well-defined vacuum state is not clear from the outset. We anticipate that if the resulting models yield the correct Lagrangian contributions for standard gauge and scalar fields, there is hope that a proper definition of a vacuum for the theory might exist.

This invariant strongly suggests the adoption of \({\hat{T}_{[o]}}\) in (\ref{Dim0End}) as the generator of a potential future internal symmetry. The Casimir (\ref{CasimirDim0}) includes a term \(M^{\rho\sigma}M_{\rho\sigma}\), which is a Casimir of the Lorentz subalgebra. There are, in fact, two quadratic Casimirs for the semisimple Lorentz subalgebra. Whether there are additional Casimirs for the found extension (\ref{Dim0Start})--(\ref{Dim0End}) of the Lorentz algebra, possibly distinguishing left-spin and right-spin representations, remains unknown.

This extension encompasses not only the internal \(U(1)\) symmetry generator \({T_{[o]}}\) but also the external Lorentz symmetry generators \(M_{\rho\sigma}\), a novel traceless symmetric tensor (spin 2) generator \(T_{\rho\sigma}\), and a would-be internal symmetry generator \({\hat{T}_{[o]}} = {T_{[o]}} + 5{\cal T}\). However, each further exploration will contribute to a deeper understanding of these extensions.

We now provide some comments on Casimirs for some of the Lorentz extension (LE) subalgebras. The 15--dimensional subalgebra \(<\{M_{\mu\nu},T_{\rho\sigma}\}>\) has a quadratic Casimir
 \bee{casdim15subalg}
 I_1=M^{\mu\nu}M_{\mu\nu}+T^{\rho\sigma}T_{\rho\sigma}.
 \ee
Another quartic Casimir is expected. There are subalgebras \(<\{M_{\mu\nu},T_{\rho\sigma}, T_\omega\}>\) and \(<\{M_{\mu\nu},T_{\rho\sigma}, \hat{T}_\omega\}>\) of dimension 19, corresponding to inhomogeneous extensions of the 15-dimensional one.

Now, the subalgebra \(<\{M_{\mu\nu},{\cal P}_{\rho}\}>\), with \({\cal P}_{\rho}\) being \((T_{\rho}+\hat{T}_{\rho})/\sqrt{2}\) or \(\ip(T_{\rho}-\hat{T}_{\rho})/\sqrt{2}\), generates a de Sitter (\(\epsilon=-1\))/Anti-de Sitter (\(\epsilon=1\)) algebra. It has a quadratic Casimir
\bee{casdim10subalg}
 I_2=\frac{1}{\epsilon}{\cal P}^{\rho}{\cal P}_{\rho}+M^{\rho\sigma}M_{\rho\sigma}.
 \ee
A quartic  casimir is obtained extrapolating on results of \cite{AyalaHasse} to include $\epsilon$:
\bee{QuarticCasdim10subalg}
I_3=\frac{1}{\epsilon}\left(M^{\rho\sigma}M_{\rho\sigma}{\cal P}^{\mu}{\cal P}_{\mu}-M^{\nu\sigma}{\cal P}_{\nu}M_{\mu\sigma}{\cal P}_{\mu}\right)+\frac{1}{4}M^{\rho\sigma}M_{\rho\sigma}M^{\mu\nu}M_{\mu\nu}.
\ee

We explore potential (pseudo-scalar) invariants involving the Levi-Civita tensor in Minkowski space. We discover the following prospective Casimir, which, as it turns out, is just an identity. We refer to it as a hollow Casimir (non-homogeneous identity):
 \bee{casdim10subalgHollow}
 I_4=\epsilon^{\mu\nu\rho\sigma}\left(\frac{1}{\epsilon^2}{\cal P}_{\mu}{\cal P}_{\nu}{\cal P}_{\rho}{\cal P}_{\sigma}+M_{\mu\nu}M_{\rho\sigma}\right)=0. 
\ee
As we will see, these invariants, although trivial like this one, could be instrumental in modeling certain portions of the Lagrangian density for the fields that embody the symmetry. The attentive reader may notice the resemblance to anomalous chiral symmetry terms. In fact, for the full Lorentz extension (LE), we can include the following hollow Casimir (non-homogeneous identity):
 \bee{casNLEHollow}
I_5=\epsilon^{\mu\nu\rho\sigma}\left({\frac{4}{\epsilon^2}}T_{\mu}\hat{T}_{\nu}T_{\rho}\hat{T}_{\sigma}+M_{\mu\nu}M_{\rho\sigma}\right)=0. 
\ee

\section{Field representation, locally invariant model}
In this section we explore the field representation of the Lorentz extension (LE) before inquiring about its representation in terms of differential operators. We develop a formalism to obtain what resembles a coadjoint representation on the fields of the considered algebra, specifically for the dimensionless subalgebra (\ref{Dim0Start})--(\ref{Dim0End}). In addition to the standard gauge theoretical representation on the fields and the construction of an invariant Lagrangian, we make considerations about possible Higgs--like fields associated with the trace parts of some fields. We also examine the possibility of expressing Lorentz symmetric and antisymmetric tensor fields in terms of vector fields without compromising unitarity.

Moving to an interaction basis with generators ${T_{[o]}}_{\text{int}}:=h{T_{[o]}}$, $ {\hat{T}_{[o]\text{int}}}:={\hat{h}}{\hat{T}_{[o]}},\ T_{\tau\ \text{int}}:=\ell\, T_{\tau} ,\ \hat{T}_{\tau\ \text{int}}:=\ell\, \hat{T}_{\tau},\ M_{\rho\sigma\ \text{int}}:={r'}M_{\rho\sigma}, T_{\rho\sigma\ \text{int}}:=r T_{\rho\sigma}$, we maintain the use of the original renormalized algebra (\ref{Dim0Start})--(\ref{Dim0End}) and keep track of the coupling constants signaling an interaction basis. We introduce several coupling constants with the expectation that, after some symmetry breaking or a similar mechanism, each generator type will evolve independently. To each dimensionless symmetry generator, we associate a field (connection) of naive dimension 1.

We will develop the covariant derivatives in a progressive fashion, starting with the standard procedure. We reserve the middle Greek letters $\varepsilon,  \kappa, \eta$ for (Einstein) space-time indices associated to covariant derivatives and later Greek letters  $\mu, \nu, \rho, \sigma, \tau, \omega, \delta$ for Lorentz indices associated with the Lorentz extension. We propose first a covariant derivative ${\cal D}_{\varepsilon}$:
\beaa
{\cal D}_{\varepsilon}:&=&\partial_{\varepsilon}+\ip h B_{\varepsilon} {T_{[o]}}+\ip \hat{h} \hat{B}_{\varepsilon} {\hat{T}_{[o]}}+ \ip \ell (\hat{\phi}_{\varepsilon}^{\, \, \tau}T_{\tau}+\phi_{ \varepsilon}^{\, \, \tau}\hat{T}_{\tau})\nn\\
& &+\ip r' B^{\, \, [\rho\sigma]}_{\varepsilon}M_{\rho\sigma}+\ip r B^{\, \, (\rho\sigma)}_{\varepsilon}T_{\rho\sigma},
\eeaa
where the fields \(B_{\varepsilon},\hat{B}_{\varepsilon},\hat{\phi}_{ \varepsilon}^{\,\, \tau},\phi_{ \varepsilon}^{\,\,\tau},B^{\,\,[\rho\sigma]}_{ \varepsilon},B^{\,\,(\rho\sigma)}_{ \varepsilon}\) are connections (gauge fields) corresponding respectively to the generators \({T_{[o]}},{\hat{T}_{[o]}},T_{\tau},\hat{T}_{\tau},M_{\rho\sigma},T_{\rho\sigma}\), with respective coupling constants \(h,\hat{h} ,\ell,\ell, r',r\).

Between the connection and the symmetry generator, we have a tensor product \(\otimes\), which we also omit to alleviate notation. In place of \(\partial_{\varepsilon}\), we would write \(\partial_{\varepsilon}\otimes Id\), with \(Id\) being the universally commuting identity operator.

For each generator \({\cal O}\) and each covariant derivative \({\cal D}_{\varepsilon}\), we define a generic quantum gauge field  \({\cal C}_{\kappa}^{\ {\cal O}}\) and the generic field strength \(\tilde{F}_{\ \varepsilon\kappa}^{{\cal O}}\):
\bee{FS}\tilde{F}_{\ \varepsilon\kappa}^{{\cal O}}:=\partial_{\varepsilon}{\cal C}_{\kappa}^{\ {\cal O}}-\partial_{\kappa}{\cal C}_{\varepsilon}^{\ {\cal O}}+{\cal F}_{\ \varepsilon\kappa}^{{\cal O}},
\ee
where ${\cal F}_{\ \varepsilon\kappa}^{{\cal O}}$ is quadratic in the fields. The formal commutator $[{\cal D}_{\varepsilon},{\cal D}_{\kappa}]$ is expanded in terms of the field strengths with the notation
\beaa
 \,[{\cal D}_{\varepsilon},{\cal D}_{\kappa}]\!\!&=&\!\!\ip (\tilde{F}_{\varepsilon\kappa}) h\, {T_{[o]}}+ \ip({\tilde{\hat{F}}}_{\varepsilon\kappa}) \hat{h}\, {\hat{T}_{[o]}}+
 \ip (\tilde{\hat{{F}}}_{\ \varepsilon\kappa}^{\rho}) \ell\, T_{\rho} +\ip(\tilde{F}_{\ \varepsilon\kappa}^{\rho}) \ell\, \hat{T}_{\rho} \nn\\
\!\!&&\!\!+ \ip(\tilde{G}_{\ \ \ \varepsilon\kappa}^{ [\tau\omega]}) r'\, M_{\tau\omega}+
 \ip(\tilde{H}_{\ \ \ \varepsilon\kappa}^{(\tau\omega)}) r \,T_{\tau\omega}.
 \eeaa
Actual computation using the interaction basis leads to quadratic parts \({\cal F}_{\ \varepsilon\kappa}^{{\cal O}}\) associated respectively with the generators ${\cal O}$ in \({T_{[o]}},{\hat{T}_{[o]}}, \hat{T}_{\rho}, T_{\rho}, M_{\rho\sigma}, T_{\rho\sigma}\):
\beaa
{\cal F}_{\varepsilon\kappa}\!\!\!&=&\!\!\!\!-\!\ip \hat{\phi}_{\varepsilon}^{\, \, \rho}\!\left(\!-\frac{12\epsilon \ell^2}{5h}g_{\rho\sigma}\!\right)\!\phi_{\kappa}^{\, \, \sigma}\!-\ip\phi_{\varepsilon}^{\, \, \rho}\!\left(\!+\frac{12\epsilon \ell^2}{5h}g_{\rho\sigma}\!\right)\!\hat{\phi}_{\kappa}^{\, \, \sigma}\!,\label{QuadraticPartInic}\\
\hat{{\cal F}}_{\varepsilon\kappa}\!\!\!&=&\!\!\!\!-\!\ip \hat{\phi}_{\varepsilon}^{\, \, \rho}\!\left(\!-\frac{\epsilon \ell^2}{10\hat{h}}g_{\rho\sigma}\!\right)\!\phi_{\kappa}^{\, \, \sigma}-\!\ip\phi_{\varepsilon}^{\, \, \rho}\!\left(\!+\frac{\epsilon \ell^2}{10\hat{h}}g_{\rho\sigma}\!\right)\!\hat{\phi}_{\kappa}^{\, \, \sigma},
\eeaa
\beaa
{\cal F}_{\ \varepsilon\kappa}^{\rho}\!\!\!&=&\!\!\!\!-\ip B_{\varepsilon}(+h)\phi_{\kappa}^{\, \, \rho}-\ip \phi_{\varepsilon}^{\, \, \rho}(-h) B_{\kappa} -\ip \hat{B}_{\varepsilon}(+\hat{h})\phi_{\kappa}^{\, \, \rho}-\ip \phi_{\varepsilon}^{\, \, \rho}(-\hat{h}) \hat{B}_{\kappa}\nn\\
&&\hspace{1.1cm} -\ip B_{\varepsilon}^{\, \, [\tau\omega]}(+\ip r'\sigma[\tau\omega]_{\sigma}^{\,\,\rho})\phi_{\kappa}^{\, \, \sigma}-\ip \phi_{\varepsilon}^{\, \, \sigma}(-\ip r'\sigma[\tau\omega]_{\sigma}^{\,\,\rho}) B_{\kappa}^{\, \, [\tau\omega]}\nn\\
&&\hspace{1.1cm} -\ip B_{\varepsilon}^{\, \, (\tau\omega)}(+ r\sigma(\tau\omega)_{\sigma}^{\,\,\rho})\phi_{\kappa}^{\, \, \sigma}-\ip \phi_{\varepsilon}^{\, \, \sigma}(- r\sigma(\tau\omega)_{\sigma}^{\,\,\rho}) B_{\kappa}^{\, \, (\tau\omega)},\\
\hat{{\cal F}}_{\ \varepsilon\kappa}^{\rho}\!\!\!&=&\!\!\!\!-\ip B_{\varepsilon}(-h)\hat{\phi}_{\kappa}^{\, \, \rho}-\ip \hat{ \phi}_{\varepsilon}^{\, \, \rho}(+h) B_{\kappa} -\ip \hat{B}_{\varepsilon}(-\hat{h})\hat{\phi}_{\kappa}^{\, \, \rho}-\ip \hat{\phi}_{\varepsilon}^{\, \, \rho}(+\hat{h}) \hat{B}_{\kappa}\nn\\
&&\hspace{1.1cm} -\ip B_{\varepsilon}^{\, \, [\tau\omega]}(+\ip r'\sigma[\tau\omega]_{\sigma}^{\,\,\rho})\hat{\phi}_{\kappa}^{\, \, \sigma}-\ip \hat{\phi}_{\varepsilon}^{\, \, \sigma}(-\ip r'\sigma[\tau\omega]_{\sigma}^{\,\,\rho}) B_{\kappa}^{\, \, [\tau\omega]}\nn\\
&&\hspace{1.1cm} -\ip B_{\varepsilon}^{\, \, (\tau\omega)}(- r\sigma(\tau\omega)_{\sigma}^{\,\,\rho})\hat{\phi}_{\kappa}^{\, \, \sigma}-\ip \hat{\phi}_{\varepsilon}^{\, \, \sigma}(+ r\sigma(\tau\omega)_{\sigma}^{\,\,\rho}) B_{\kappa}^{\, \, (\tau\omega)},\\
{\cal G}_{\ \ \ \varepsilon\kappa}^{ [\tau\omega]}\!\!\!&=&\!\!\!\!
-\!\ip \hat{\phi}_{\varepsilon}^{\, \, \mu}\!\!\left(\!+\frac{\ip \epsilon \ell^2}{2r'}(\delta_{\mu}^{\,\,\tau}\delta_{\nu}^{\,\,\omega}\!-\!\delta_{\mu}^{\,\,\omega}\delta_{\nu}^{\,\,\tau})\!\right)\!\!\phi_{\kappa}^{\, \, \nu}\!-\!\ip\phi_{\varepsilon}^{\, \, \mu}\!\!\left(\!+\frac{\ip \epsilon \ell^2}{2r'}(\delta_{\mu}^{\,\,\tau}\delta_{\nu}^{\,\,\omega}\!-\!\delta_{\mu}^{\,\,\omega}\delta_{\nu}^{\,\,\tau})\!\right)\!\!\hat{\phi}_{\kappa}^{\, \, \nu}\!\!\nn\\
&&\hspace{1.1
	cm} -\ip B_{\varepsilon}^{\, \, [\rho\sigma]}(+\ip r'\left(\sigma[\rho\sigma]_{\mu}^{\,\,\tau}\delta_{\nu}^{\,\,\omega}+\sigma[\rho\sigma]_{\nu}^{\,\,\omega}\delta_{\mu}^{\,\,\tau}\right))B_{\kappa}^{\, \, [\mu\nu]}\nn\\
 &&\hspace{1.1
 	cm} -\ip B_{\varepsilon}^{\, \, (\rho\sigma)}\left(+ \frac{\ip r^2}{r'}\left(\sigma(\rho\sigma)_{\mu}^{\,\,\tau}\delta_{\nu}^{\,\,\omega}-\sigma(\rho\sigma)_{\nu}^{\,\,\omega}\delta_{\mu}^{\,\,\tau}\right)\right) B_{\kappa}^{\, \, (\mu\nu)},\\\
 {\cal H}_{\ \ \ \varepsilon\kappa}^{(\tau\omega)}\!\!\!&=&\!\!\!\!
 -\ip B_{\varepsilon}^{\, \, [\rho\sigma]}(+\ip r'\left(\sigma[\rho\sigma]_{\mu}^{\,\,\tau}\delta_{\nu}^{\,\,\omega}+\sigma[\rho\sigma]_{\nu}^{\,\,\omega}\delta_{\mu}^{\,\,\tau}\right))B_{\kappa}^{\, \, (\mu\nu)}\nn\\
 &&\hspace{1.1
 	cm} -\ip B_{\varepsilon}^{\, \, (\mu\nu)}(-\ip r'\left(\sigma(\mu\nu)_{\rho}^{\,\,\tau}\delta_{\sigma}^{\,\,\omega}-\sigma(\mu\nu)_{\sigma}^{\,\,\omega}\delta_{\rho}^{\,\,\tau}\right)) B_{\kappa}^{\, \,[\rho\sigma]}\nn\\
 &&\hspace{1.1
 	cm}-\!\ip \hat{\phi}_{\varepsilon}^{\, \, \mu}\!\left(\!-\frac{ \epsilon \ell^2}{2r}\left(\delta_{\mu}^{\,\,\tau}\delta_{\nu}^{\,\,\omega}+\delta_{\mu}^{\,\,\omega}\delta_{\nu}^{\,\,\tau}-\half g^{\tau\omega}g_{\mu\nu}\right)\!\right)\!\phi_{\kappa}^{\, \, \nu}\nn\\
 && \hspace{1.1
 	cm}-\ip\phi_{\varepsilon}^{\, \, \mu}\!\left(\!+\frac{ \epsilon \ell^2}{2r}\left(\delta_{\mu}^{\,\,\tau}\delta_{\nu}^{\,\,\omega}+\delta_{\mu}^{\,\,\omega}\delta_{\nu}^{\,\,\tau}-\half g^{\tau\omega}g_{\mu\nu}\right)\!\right)\!\hat{\phi}_{\kappa}^{\, \, \nu}.\label{QuadraticPartEnd} 
\eeaa
We adopt thus the generic Ansatz 
 \beaa
\!\lambda^{{\cal O}}{\cal F}_{\ \varepsilon\kappa}^{{\cal O}}\!=\!-{\ip} \sum_{{\cal O}''}\!\!\left({\cal C}_{\varepsilon}^{\ {\cal O}''}\lambda^{{\cal O}''}[{\cal O}'',\lambda^{{\cal O}}{\cal C}_{\kappa}^{\ {\cal O}}]_1 \right)\!=\!{\ip} (\lambda^{{\cal O}''}{\cal C}_{\varepsilon}^{\ {\cal O}''})\ip C_{\ {\cal O}''{\cal O}'}^{{\cal O}}(\lambda^{{\cal O}'}{\cal C}_{\varepsilon}^{\ {\cal O}'}),\label{FieldStrength}&&\\
\,[{\cal O}'',\lambda^{{\cal O}}{\cal C}_{\varepsilon}^{\ {\cal O}}]_1\!=\!- \ip C_{\ {\cal O}''{\cal O}'}^{{\cal O}}(\lambda^{{\cal O}'}{\cal C}_{\varepsilon}^{\ {\cal O}'}),\ {\rm for}\ \ [{\cal O}'',{\cal O}']=\ip\,C_{\ {\cal O}''{\cal O}'}^{{\cal O}}\,{\cal O}.&&
\eeaa   
From it, we can read the field  transformations, for $b=\varepsilon\in\{0,1,2,3\}$:
\beaa
\,[ T_{\mu},hB_{b}]_1&=&\!\left(\!-\frac{12\epsilon \ell}{5}g_{\mu\sigma}\!\right)\!\phi_{b}^{\, \, \sigma},\hspace{.3cm}[\hat{T}_{\mu},hB_{b}]_1\ =\ \!\left(\!+\frac{12\epsilon \ell}{5}g_{\mu\sigma}\!\right)\!\hat{\phi}_{b}^{\, \, \sigma},\label{Eqn27}\\
\,[ T_{\mu},\hat{h}\hat{B}_{b}]_1&=&\!\left(\!-\frac{\epsilon \ell}{10}g_{\mu\sigma}\!\right)\!\phi_{b}^{\, \, \sigma},\hspace{.6cm}[\hat{T}_{\mu},\hat{h}\hat{B}_{b}]_1\ =\ \!\left(\!+\frac{\epsilon \ell}{10}g_{\mu\sigma}\!\right)\!\hat{\phi}_{b}^{\, \, \sigma},\label{Eqn28}\\
\,[T_{\mu},r'B_{b}^{\, \, [\tau\kappa]}]_1&=&\!\left(\!+\frac{\ip \epsilon \ell}{2}(\delta_{\mu}^{\,\,\tau}\delta_{\delta}^{\,\,\kappa}-\delta_{\mu}^{\,\,\kappa}\delta_{\delta}^{\,\,\tau})\!\right)\!\phi_{b}^{\, \, \delta},\nn\\
&&\!\!\![\hat{T}_{\mu},r'B_{b}^{\, \, [\tau\kappa]}]_1\ =\ \!\!\left(\!+\frac{\ip \epsilon \ell}{2}(\delta_{\mu}^{\,\,\tau}\delta_{\delta}^{\,\,\kappa}-\delta_{\mu}^{\,\,\kappa}\delta_{\delta}^{\,\,\tau})\!\right)\!\hat{\phi}_{b}^{\, \, \delta},\\
%\eeaa
%\beaa
\,[T_{\mu},rB_{b}^{\, \, (\tau\kappa)}]_1&=&\!\left(\!-\frac{\epsilon \ell}{2}(\delta_{\mu}^{\,\,\tau}\delta_{\delta}^{\,\,\kappa}+\delta_{\mu}^{\,\,\kappa}\delta_{\delta}^{\,\,\tau}\!-\half g^{\tau\kappa}g_{\mu\delta})\!\right)\!\phi_{b}^{\, \, \delta},\nn\\
&&\!\!\![\hat{T}_{\mu},rB_{b}^{\, \, (\tau\kappa)}]_1=\ \!\!\left(\!\frac{ \epsilon \ell}{2}(\delta_{\mu}^{\,\,\tau}\delta_{\delta}^{\,\,\kappa}\!+\!\delta_{\mu}^{\,\,\kappa}\delta_{\delta}^{\,\,\tau}\!-\!\half g^{\tau\kappa}g_{\mu\delta})\!\right)\!\hat{\phi}_{b}^{\, \, \delta},\ \ \ \\
\,[T_{\mu},\ell\,\phi_{b}^{\, \, \tau}]&=&0,\hspace{2.9cm}[\hat{T}_{\mu},\ell\,\hat{\phi}_{b}^{\, \, \tau}]_1\ =\ 0,
\eeaa
\beaa
\,[T_{\mu},\ell\,\hat{\phi}_{b}^{\, \, \tau}]_1&=&\!h \delta_{\mu}^{\,\,\tau} B_{b}+\hat{h} \delta_{\mu}^{\,\,\tau} \hat{B}_{b}-\!\ip r'\sigma[\rho\sigma]_{\mu}^{\,\,\tau}B_{b}^{\, \, [\rho\sigma]}+r\sigma(\rho\sigma)_{\mu}^{\,\,\tau}\!B_{b}^{\, \, (\rho\sigma)}\!\!\!,\nn\\
\,[\hat{T}_{\mu},\ell\,\phi_{b}^{\, \, \tau}]_1&=&\!\!\!\!\!-h \delta_{\mu}^{\,\,\tau} B_{b}-\hat{h} \delta_{\mu}^{\,\,\tau} \hat{B}_{b}-\!\ip r'\sigma[\rho\sigma]_{\mu}^{\,\,\tau}B_{b}^{\, \, [\rho\sigma]}-r\sigma(\rho\sigma)_{\mu}^{\,\,\tau}\!B_{b}^{\, \, (\rho\sigma)}\!\!\!,\ \ \ \label{Eqn32}\\
\,[{T_{[o]}},hB_{b}]_1&=&0,\hspace{2.9cm}[{T_{[o]}},\hat{h}\hat{B}_{b}]_1\ =\ 0,\\
\,[{T_{[o]}},r'B_{b}^{\, \, [\tau\kappa]}]_1&=&0,\hspace{2.4cm}[{T_{[o]}},rB_{b}^{\, \, (\tau\kappa)}]_1\ =\ 0,\\
\,[{T_{[o]}},\ell\,\phi_{b}^{\, \, \tau}]_1&=&+\ell\,\phi_{b}^{\, \, \tau},\hspace{2.0cm}[{T_{[o]}},\ell\,\hat{\phi}_{b}^{\, \, \tau}]_1\ =-\ell\,\hat{\phi}_{b}^{\, \, \tau},\\
\,[{\hat{T}_{[o]}},hB_{b}]_1&=&0,\hspace{2.9cm}[{\hat{T}_{[o]}},\hat{h}\,\hat{B}_{b}]_1\ =\ 0,\\
\!\!\![{\hat{T}_{[o]}},r'B_{b}^{\, \, [\tau\kappa]}]_1&=&0,\hspace{2.4cm}[{\hat{T}_{[o]}},rB_{b}^{\, \, (\tau\kappa)}]_1\ =\ 0,\\
\,[{\hat{T}_{[o]}},\ell\,\phi_{b}^{\, \, \tau}]_1&=&+\ell\,\phi_{b}^{\, \, \tau},\hspace{2.0cm}[{\hat{T}_{[o]}},\ell\,\hat{\phi}_{b}^{\, \, \tau}]_1\ =-\ell\,\hat{\phi}_{b}^{\, \, \tau},\\
\,[M_{\rho\sigma},hB_{b}]_1&=&0,\hspace{2.9cm}[ M_{\rho\sigma},\hat{h}\hat{B}_{b}]_1 =\ 0,\\
\!\!\![ M_{\rho\sigma},r'B_{b}^{\, \, [\tau\kappa]}]_1&=&+\ip r'(\sigma[\rho\sigma]_{\gamma}^{\,\,\tau}\delta_{\delta}^{\,\,\kappa}+\sigma[\rho\sigma]_{\delta}^{\,\,\kappa}\delta_{\gamma}^{\tau})B_{b}^{\, \, [\gamma\delta]},\\
\!\!\![ M_{\rho\sigma},rB_{b}^{\, \, (\tau\kappa)}]_1&=&+\ip r(\sigma[\rho\sigma]_{\gamma}^{\,\,\tau}\delta_{\delta}^{\,\,\kappa}+\sigma[\rho\sigma]_{\delta}^{\,\,\kappa}\delta_{\gamma}^{\tau})B_{b}^{\, \, (\gamma\delta)},\\
\,[ M_{\rho\sigma},\ell\,\phi_{b}^{\, \, \tau}]_1&=&+\ip \ell\,\sigma[\rho\sigma]_{\gamma}^{\,\,\tau}\phi_{b}^{\, \, \gamma},\hspace{.2cm}
[ M_{\rho\sigma},\ell\,\hat{\phi}_{b}^{\, \, \tau}]_1\ =\ +\ip \ell\,\sigma[\rho\sigma]_{\gamma}^{\,\,\tau}\hat{\phi}_{b}^{\, \, \gamma},\ 
\eeaa
\beaa
\!\![ T_{\rho\sigma},hB_{b}]_1&=&0,\hspace{2.9cm}[T_{\rho\sigma},\hat{h}\hat{B}_{b}]_1 =\ 0,\\
\,[T_{\rho\sigma},r'B_{b}^{\, \, [\tau\kappa]}]_1&=&+\ip r(\sigma(\rho\sigma)_{\gamma}^{\,\,\tau}\delta_{\delta}^{\,\,\kappa}-\sigma(\rho\sigma)_{\delta}^{\,\,\kappa}\delta_{\gamma}^{\,\,\tau})B_{b}^{\, \, (\gamma\delta)},\\
\!\![T_{\rho\sigma},rB_{b}^{\, \, (\tau\kappa)}]_1&=&+\ip r'(\sigma(\rho\sigma)_{\gamma}^{\,\,\tau}\delta_{\delta}^{\,\,\kappa}-\sigma(\rho\sigma)_{\delta}^{\,\,\kappa}\delta_{\gamma}^{\,\,\tau})B_{b}^{\, \, [\gamma\delta]},\\
\,[ T_{\rho\sigma},\ell\,\phi_{b}^{\, \, \tau}]_1&=&+ \ell\,\sigma(\rho\sigma)_{\gamma}^{\,\,\tau}\phi_{b}^{\, \, \gamma},\hspace{.6cm}
[ T_{\rho\sigma},\ell\,\hat{\phi}_{b}^{\, \, \tau}]_1\ =\  -\ell\,\sigma(\rho\sigma)_{\gamma}^{\,\,\tau}\hat{\phi}_{b}^{\, \, \gamma}.\label{Eqn46}
\eeaa
It is easy to verify that the connections \(hB_{\kappa},\hat{h}\hat{B}_{\kappa},\ell\,\hat{\phi}_{ \kappa}^{\,\, \tau},\ell\,\phi_{ \kappa}^{\,\,\tau},r'B^{\,\,[\tau\omega]}_{ \kappa},rB^{\,\,(\tau\omega)}_{ \kappa}\) with \([{\cal O},\cdot]_1\) build a left Lie algebra module, that is, it is a representation of the LE algebra (\ref{Dim0Start})--(\ref{Dim0End}).  Notice that the subindex \(b=\kappa\) is an Einstein space-time label which is not transformed, nor affected, by the considered transformations. This subindex is attached to the local space-time coordinate derivative which  is assumed thus far to commute with all symmetry transformations. This is obviously not necessarily the case. Further elaborations will address the transformation of geometric or coordinate space--time frame in a further section.

The first--order gauge transformations $({\cal C}_{\mu}^{\ {\cal O}})^\prime$ with space--time point-dependent parameters $\varsigma(x),\hat{\varsigma}(x)$, $\varsigma^{[\tau\omega]}(x),\varsigma^{(\tau\omega)}(x),\hat{\varsigma}^{\nu}(x),\varsigma^{\nu}(x)$ associated respectively with transformations with generators ${T_{[o]}}, \bar{T}_{[o]},M_{\tau\omega},T_{\tau\omega},T_{\nu},\hat{T}_{\nu}$ are given by:
\beaa
(B_{\kappa})'&=& B_{\kappa}+\ip\frac{\hat{\varsigma}^{\nu}(x)}{\ell}\left(\!-\frac{12\epsilon \ell^2}{5h}g_{\nu\sigma}\!\right)\!\phi_{\kappa}^{\, \, \sigma}\nn\\
&&\hspace{.4cm}+\ip\frac{\varsigma^{\nu}(x)}{\ell}\left(\!+\frac{12\epsilon \ell^2}{5h}g_{\nu\sigma}\!\right)\!\hat{\phi}_{\kappa}^{\, \, \sigma}+\frac{1}{h}\partial_{\kappa}\varsigma(x),\label{FOTInic}\\
(\hat{B}_{\kappa})'&=& \hat{B}_{\kappa}+\ip\frac{\hat{\varsigma}^{\nu}(x)}{\ell}\left(\!-\frac{\epsilon \ell^2}{10\hat{h}}g_{\nu\sigma}\!\right)\!\phi_{\kappa}^{\, \, \sigma}\nn\\
&&\hspace{.4cm}+\ip\frac{\varsigma^{\nu}(x)}{\ell}\left(\!+\frac{\epsilon \ell^2}{10\hat{h}}g_{\nu\sigma}\!\right)\!\hat{\phi}_{\kappa}^{\, \, \sigma}+\frac{1}{\hat{h}}\partial_{\kappa}\hat{\varsigma}(x),\\
(\phi_{ \kappa}^{\, \, \rho})'\!\!\!\!&=&\!\!\!\!\phi_{ \kappa}^{\, \, \rho}
+\ip \frac{ \varsigma^{\sigma}(x)}{\ell}\!\left(\!
-h\delta_{\sigma}^{\,\,\rho} B_{\kappa}\!
-\hat{h}\delta_{\sigma}^{\,\,\rho}\hat{B}_{\kappa}\!
-\!\ip r'\sigma[\tau\omega]_{\sigma}^{\,\,\rho} B_{\kappa}^{\, \, [\tau\omega]} 
- r\sigma(\tau\omega)_{\sigma}^{\,\,\rho} B_{\kappa}^{\, \, (\tau\omega)}\!\right)\nn\\
&&+\ip \frac{\varsigma^{ [\tau\omega]}(x)}{r'}(+\ip r'\sigma[\tau\omega]_{\sigma}^{\,\,\rho})\phi_{\kappa}^{\, \, \sigma}
+\ip \frac{\varsigma^{ (\tau\omega)}(x)}{r}(+ r\sigma(\tau\omega)_{\sigma}^{\,\,\rho})\phi_{\kappa}^{\, \, \sigma}\nn\\
&&   +\ip  \frac{\varsigma(x)}{h}(+h)\phi_{\kappa}^{\, \, \rho}
+\ip \frac{\hat{\varsigma}(x)}{\hat{h}}(+\hat{h})\phi_{\kappa}^{\, \, \rho}\!+\!\frac{1}{\ell}\partial_{\kappa}\varsigma^{\rho}(x),\\
(\hat{\phi}_{ \kappa}^{\, \, \rho})'\!\!\!\!&=&\!\!\!\!\hat{\phi}_{ \kappa}^{\, \, \rho}
+\ip \frac{
	 \hat{\varsigma}^{\sigma}(x)}{\ell}\!\left(\!
h\delta_{\sigma}^{\,\,\rho} B_{\kappa}\!
+\hat{h}\delta_{\sigma}^{\,\,\rho} \hat{B}_{\kappa}\!
-\!\ip r'\sigma[\tau\omega]_{\sigma}^{\,\,\rho} B_{\kappa}^{\, \, [\tau\omega]} 
+ r\sigma(\tau\omega)_{\sigma}^{\,\,\rho} B_{\kappa}^{\, \, (\tau\omega)}\!\right)\nn\\
&&+\ip \frac{\varsigma^{ [\tau\omega]}(x)}{r'}(+\ip r'\sigma[\tau\omega]_{\sigma}^{\,\,\rho})\hat{\phi}_{\kappa}^{\, \, \sigma}
+\ip \frac{\varsigma^{ (\tau\omega)}(x)}{r}(- r\sigma(\tau\omega)_{\sigma}^{\,\,\rho})\hat{\phi}_{\kappa}^{\, \, \sigma}\nn\\
&&   +\ip  \frac{\varsigma(x)}{h}(-h)\hat{\phi}_{\kappa}^{\, \, \rho}
+\ip \frac{\hat{\varsigma}(x)}{\hat{h}}(-\hat{h})\hat{\phi}_{\kappa}^{\, \, \rho}\!+\!\frac{1}{\ell}\partial_{\kappa} \hat{\varsigma}^{\rho}(x),
\eeaa
\beaa
\!\!\!(B_{\kappa}^{\, \, [\tau\omega]})'\!\!\!\!&=&\!\!\!\!B_{\kappa}^{\, \, [\tau\omega]}
+\ip \frac{\varsigma^{ [\rho\sigma]}(x)}{r'}(+\ip r'\left(\sigma[\rho\sigma]_{\mu}^{\,\,\tau}\delta_{\nu}^{\,\,\omega}+\sigma[\rho\sigma]_{\nu}^{\,\,\omega}\delta_{\mu}^{\,\,\tau}\right))B_{\kappa}^{\, \, [\mu\nu]}\!\!\!\nn\\
&&\hspace{0.4
	cm} +\ip \frac{{\varsigma}^{(\rho\sigma)}(x)}{r}\left(- \frac{\ip r^2}{r'}\left(\sigma(\rho\sigma)_{\mu}^{\,\,\tau}\delta_{\nu}^{\,\,\omega}-\sigma(\rho\sigma)_{\nu}^{\,\,\omega}\delta_{\mu}^{\,\,\tau}\right)\right) B_{\kappa}^{\, \, (\mu\nu)}\!\!\!\nn\\
&&\hspace{0.4
	cm}+\!\ip \frac{\hat{\varsigma}^{ \mu}(x)}{\ell}\!\left(\!+\frac{\ip \epsilon \ell^2}{2r'}(\delta_{\mu}^{\,\,\tau}\delta_{\nu}^{\,\,\omega}-\delta_{\mu}^{\,\,\omega}\delta_{\nu}^{\,\,\tau})\!\right)\!\phi_{\kappa}^{\, \, \nu}\!\!\!\nn\\
&& \hspace{0.4
	cm}+\ip\frac{\varsigma^{ \mu}(x)}{\ell}\!\left(\!+\frac{\ip \epsilon \ell^2}{2r'}(\delta_{\mu}^{\,\,\tau}\delta_{\nu}^{\,\,\omega}-\delta_{\mu}^{\,\,\omega}\delta_{\nu}^{\,\,\tau})\!\right)\!\hat{\phi}_{\kappa}^{\, \, \nu}+\frac{1}{r'}\partial_{\kappa}\varsigma^{ [\tau\omega]}(x), 
\!\!\!
\eeaa
\beaa
(B_{\kappa}^{\, \, (\tau\omega)})'\!\!\!\!&=&\!\!\!\!B_{\kappa}^{\, \, (\tau\omega)}
+\ip  \frac{\varsigma^{ [\rho\sigma]}(x)}{r'}(+\ip r'\left(\sigma[\rho\sigma]_{\mu}^{\,\,\tau}\delta_{\nu}^{\,\,\omega}+\sigma[\rho\sigma]_{\nu}^{\,\,\omega}\delta_{\mu}^{\,\,\tau}\right))B_{\kappa}^{\, \, (\mu\nu)}\nn\\
&&\hspace{0.4
	cm} +\ip\frac{{\varsigma}^{(\rho\sigma)}(x)}{r}(-\ip r'\left(\sigma[\mu\nu]_{\rho}^{\,\,\tau}\delta_{\sigma}^{\,\,\omega}+\sigma[\mu\nu]_{\sigma}^{\,\,\omega}\delta_{\rho}^{\,\,\tau}\right)) B_{\kappa}^{\, \, [\mu\nu]}\nn\\
&&\hspace{0.4
	cm}+\ip \frac{\hat{\varsigma}^{ \mu}(x)}{\ell}\!\left(\!-\frac{ \epsilon \ell^2}{2r}\left(\delta_{\mu}^{\,\,\tau}\delta_{\nu}^{\,\,\omega}\!+\!\delta_{\mu}^{\,\,\omega}\delta_{\nu}^{\,\,\tau}\!-\!\half g^{\tau\omega}g_{\mu\nu}\right)\!\right)\!\phi_{\kappa}^{\, \, \nu}\nn\\
&\!\!&\!\!\!+\ip\frac{\varsigma^{ \mu}(x)}{\ell}\!\left(\!+\!\frac{ \epsilon \ell^2}{2r}\left(\delta_{\mu}^{\,\,\tau}\delta_{\nu}^{\,\,\omega}\!\!+\!\delta_{\mu}^{\,\,\omega}\delta_{\nu}^{\,\,\tau}\!-\!\half g^{\tau\omega}g_{\mu\nu}\right)\!\right)\!\hat{\phi}_{\kappa}^{\, \, \nu}\!+\!\frac{1}{r}\partial_{\kappa}\varsigma^{ (\tau\omega)}\!(x).\ \ \label{FOTEnd}
\eeaa
As in any standard gauge theory, the field strengths (curvatures) transform identically as with rigid transformations (not space--time point-dependent parameters). Calling $(\tilde{F}_{\ \varepsilon\kappa}^{{\cal O}})^\prime$ the transformed field strength, we have, for instance:
\beaa
({\tilde{ F}}_{\varepsilon\kappa})^\prime&=&{\tilde{ F}}_{\varepsilon\kappa}+\ip\frac{\hat{\varsigma}^{\omega}(x)}{\ell}\left(\!-\frac{12\epsilon \ell^2}{5h}g_{\omega\sigma}\!\right)\!{\tilde{ F}}_{\ \varepsilon\kappa}^{\sigma}\nn\\
&&\hspace{.3cm}+\ip\frac{\varsigma^{\omega}(x)}{\ell}\left(\!+\frac{12\epsilon \ell^2}{5h}g_{\omega\sigma}\!\right)\!\tilde{{\hat{ F}}}_{\ \varepsilon\kappa}^{\sigma},\\
(\tilde{{\hat{ F}}}_{\  \varepsilon\kappa}^{\rho})'&=&\tilde{{\hat{ F}}}_{\  \varepsilon\kappa}^{\rho}
+\ip  \frac{\varsigma(x)}{h}(-h)\tilde{{\hat{ F}}}_{\ \varepsilon\kappa}^{\rho}
+\ip \frac{\hat{\varsigma}(x)}{\hat{h}}(-\hat{h})\tilde{{\hat{ F}}}_{\ \varepsilon\kappa}^{\rho}\nn\\
&+&\!\!\!\!\ip \frac{ \hat{\varsigma}^{\sigma}(x)}{\ell}\!\left(\!
h\delta_{\sigma}^{\,\,\rho} {\tilde{ F}}_{\varepsilon\kappa}\!
+\hat{h}\delta_{\sigma}^{\,\,\rho} \tilde{\hat{ F}}_{\varepsilon\kappa}\!
-\!\ip r'\sigma[\tau\omega]_{\sigma}^{\,\,\rho} {\tilde{ G}}_{\ \ \ \varepsilon\kappa}^{[\tau\omega]} 
+ r\sigma(\tau\omega)_{\sigma}^{\,\,\rho} {\tilde{ H}}_{\ \ \ \varepsilon\kappa}^{(\tau\omega)}\!\right)\ \hspace{-1cm}\nn\\
&+&\!\!\!\!\ip \frac{\varsigma^{ [\tau\omega]}(x)}{r'}(+\ip r'\sigma[\tau\omega]_{\sigma}^{\,\,\rho})\tilde{{\hat{ F}}}_{\ \varepsilon\kappa}^{\sigma}
+\ip \frac{\varsigma^{ (\tau\omega)}(x)}{r}(- r\sigma(\tau\omega)_{\sigma}^{\,\,\rho})\tilde{{\hat{ F}}}_{\ \varepsilon\kappa}^{\sigma},
\eeaa  
and similar expressions for the other field strengths. The action of generators on the field strength goes along the corresponding expressions for the fields (interchanging $ B_{\varepsilon}, \hat{B}_{\varepsilon},\cdots,\hat{\phi}_{ \varepsilon}^{\, \, \rho}$ respectively with $ \tilde{ F}_{\varepsilon\kappa}, \tilde{\hat{F}}_{\varepsilon\kappa},\cdots,\tilde{\hat{F}}_{\  \varepsilon\kappa}^{\rho}$ ). For instance,
\beaa
\,[T_{\mu},\ell\,\tilde{{\hat{ F}}}_{\ \varepsilon\kappa}^{\tau}]_1\!\!\!\!&=\!&\!\!h \delta_{\mu}^{\,\,\tau} {\tilde{ F}}_{\varepsilon\kappa}+\hat{h} \delta_{\mu}^{\,\,\tau} \tilde{{\hat{ F}}}_{\varepsilon\kappa}-\!\ip r'\sigma[\rho\sigma]_{\mu}^{\,\,\tau}{\tilde{ G}}_{\ \ \  \varepsilon\kappa}^{ [\rho\sigma]}+r\sigma(\rho\sigma)_{\mu}^{\,\,\tau}\!{\tilde{H}}_{\ \ \ \varepsilon\kappa}^{(\rho\sigma)}\!\!\!,\ \ \nn\\
\,[\bar{T}_{\mu},\ell\,{\tilde{ F}}_{\ \varepsilon\kappa}^{\tau}]_1\!\!\!\!&=\!\!&\!\!\!\!-h \delta_{\mu}^{\,\,\tau} {\tilde{ F}}_{\varepsilon\kappa}-\hat{h} \delta_{\mu}^{\,\,\tau} \tilde{\hat{ F}}_{\varepsilon\kappa}-\!\ip r'\sigma[\rho\sigma]_{\mu}^{\,\,\tau}{\tilde{ G}}_{\ \ \ \varepsilon\kappa}^{ [\rho\sigma]}-r\sigma(\rho\sigma)_{\mu}^{\,\,\tau}\!{\tilde{ H}}_{\ \ \ \varepsilon\kappa}^{(\rho\sigma)}.\ \ \ 
\eeaa
Now, we would like to construct an invariant Lagrangian associated to the considered field strengths (curvatures). We use the invariant quadratic casimir $I$ in (\ref{CasimirDim0}) as our prime tool. This invariant can be rewritten as follows:
\beaa
I=\left[\frac{{T}^{\omega}}{\ell}\ \frac{\hat{T}^{\omega}}{\ell}\ \frac{M^{\rho\tau}}{r'}\ \frac{T ^{\rho\tau}}{r}\ \frac{{T_{[o]}}}{h}\ \frac{{\hat{T}_{[o]}}}{\hat{h}}\right]\!\cdot\! {\bf M}\!\cdot\! \left[ \frac{\hat{T}_{\omega}}{\ell}\ \frac{{T}_{\omega}}{\ell}\ \frac{M_{\rho\tau}}{r'}\ \frac{T_{\rho\tau}}{r}\ \frac{{T_{[o]}}}{h}\ \frac{{\hat{T}_{[o]}}}{\hat{h}}\right]^{tr}\!\!\!,\ \ &&
\\ 
{\rm with}\ \   {\bf M}:={\rm diag}\left( \frac{2}{\epsilon},\ \frac{2}{\epsilon},\ \frac{1}{1},\ 
\frac{1}{1},\ \frac{24}{5},\ \frac{1}{5}\right).\label{CasimirDim0Int}&&
\eeaa
We consider the field strength dual counterpart to build a Lagrangian \({\cal L}_o\). It is easy to verify the invariance of the Lagrangian \({\cal L}_o\) under the considered local transformations. For instance, \([T_{\mu},{\cal L}_o]=0\). This is equivalent to taking the trace of the matrix products in the adjoint representation:
\beaa
4{\cal L}_o&=&\ip[\ell\tilde{{\hat{F}}}^{\ \varepsilon\kappa}_{\omega}\ \ell{\tilde{F}}^{\ \varepsilon\kappa}_{\omega}\ r'{\tilde{G}}^{\ \ \ \varepsilon\kappa}_{[\rho\tau]}\ r{\tilde{H}}^{\ \ \ \varepsilon\kappa} _{(\rho\tau)}\ h{\tilde{F}}^{\varepsilon\kappa}\ \hat{h}\tilde{{\hat{F}}}^{\varepsilon\kappa}]\nn\\
&&\cdot  {\bf M}^{-1}\cdot \ip[\ell{\tilde{F}}_{\ \varepsilon\kappa}^{\omega}\ \ell\tilde{{\hat{F}}}_{\ \varepsilon\kappa}^{\omega}\ r'{\tilde{G}}_{\ \ \ \varepsilon\kappa}^{[\rho\tau]}\ r{\tilde{H}}_{\ \ \ \varepsilon\kappa}^{(\rho\tau)}\ h{\tilde{F}}_{\varepsilon\kappa}\ \hat{h}\tilde{{\hat{F}}}_{\varepsilon\kappa}]^{tr}\!\!\ \ \\
&=& -\frac{\epsilon\ell^2}{2}\{ \tilde{{\hat{F}}}^{\ \varepsilon\kappa}_{\omega},{\tilde{F}}_{\  \varepsilon\kappa}^{\omega}\}-{r'^2}{\tilde{G}}^{\ \ \ \varepsilon\kappa}_{ [\rho\tau]}{\tilde{G}}_{\ \ \ \varepsilon\kappa}^{[\rho\tau]}- 
{r^2}{\tilde{H}}^{\ \ \ \varepsilon\kappa} _{ (\rho\tau)}{\tilde{H}}_{\ \ \ \varepsilon\kappa}^{ (\rho\tau)}\nn\\
&&- \frac{5h^2}{24}{\tilde{F}}^{\varepsilon\kappa}{\tilde{F}}_{\varepsilon\kappa}-   {5\hat{h}^2}\tilde{{\hat{F}}}^{\varepsilon\kappa}\tilde{\hat{F}}_{\varepsilon\kappa}.
\eeaa

We briefly consider the possibility of expressing the gauge fields \(\phi_{ \varepsilon}^{\,\,\sigma}\) and \(\bar{\phi}_{ \varepsilon}^{\,\,\sigma}\) (coming from the representation \((\frac{1}{2},\frac{1}{2})\otimes(\frac{1}{2},\frac{1}{2})\) of the Lorentz-index vector label of the generator \(T_\sigma\) or \(\bar{T}_{\sigma}\) and the Einstein-index vector label of the covariant derivative) in terms of their trace through a complex scalar field \(\phi\) and potentially further contributions (likely in the form of an antisymmetric vector or a spin-\((1,1)\) symmetric tensor), setting all other gauge fields to zero. We obtain:
\beaa
\phi_{ \varepsilon}^{\,\,\sigma}\!\!&\!=\!&\!\frac{\delta_{ \varepsilon}^{\,\,\sigma}\sqrt{2}}{\sqrt{3|\epsilon|}\ell}\phi + \cdots,\label{Higgs1}\\
\!\!\!{\cal L}_o|_{restr}\!\!\!&\!\!=\!\!&\!\!\! -\frac{\epsilon\ell^2}{8}\{ {\tilde{\hat{F}}}^{\ \varepsilon\kappa}_{ \omega}\!,{\tilde{F}}_{\ \varepsilon\kappa}^{ \omega}\}-\!{\frac{r'^2}{4}}{\tilde{G}}^{\ \ \ \varepsilon\kappa}_{ [\rho\tau]}{\tilde{G}}_{\ \ \ \varepsilon\kappa}^{ [\rho\tau]}\!=\!-\frac{\epsilon}{|\epsilon|}\partial_{\varepsilon}\!\phi \partial^{\varepsilon}\bar{\phi}-\!\frac{8}{3}{(\phi\bar{\phi})}^2\!\!+\!\cdots\!.\ \ \  \label{Higgs}
\eeaa
This would require \(\epsilon\) to be negative. An additional quadratic soft-breaking mass term might complete a Higgs-like potential. The quartic self-interaction of the would-be Higgs field comes from \({\tilde{G}}_{\ \ \varepsilon\kappa}^{[\rho\tau]}\), responsible for local Lorentz transformation invariance, although with a substantial coupling constant. 

The corresponding versions of (\ref{Eqn27}) and (\ref{Eqn32}) would embody the aimed mixture between gauge and Higgs field. Nevertheless, although this field might contribute to the Higgs field perhaps after some symmetry breaking, there are fields that will properly embody the Higgs, as we will see in a Poincaré-like extension \cite{U1PoinExt}. Beware: in using \(\delta_{ \varepsilon}^{\,\,\sigma}\) in (\ref{Higgs1}), we identify an Einstein (non-Lorentz transforming) index \(\varepsilon\) with a Lorentz-transforming index \(\sigma\). In fact, the transformations (\ref{Eqn27}--\ref{Eqn28}) and (\ref{Eqn32}) represent rather a mixing between the gauge fields \(hB_b\), \(\hat{h}\hat{B}_b\) and the tetrad-like fields \(\ell\phi_b^{\,\tau}\), \(\ell\hat{\phi}_b^{\,\tau}\). The transformation between Higgs-like fields demands either a wider realm of covariant derivatives (explored below) or a Poincaré-like extension.

The kinetic parts of the gauge fields \(hB_b\), \(\hat{h}\hat{B}_b\) will give the proper sign for kinetic terms (where "tdt." stands for total derivative terms):
\beaa
-\frac{5h^2}{24\cdot 4}(\partial_{\varepsilon}B_{\kappa}\!-\partial_{\kappa}B_{\varepsilon})(\partial^{\varepsilon}B^{\kappa}\!-\partial^{\kappa}B^{\varepsilon})=\frac{5h^2}{24\cdot 4}(2B_{\kappa}(\partial_{\varepsilon}\partial^{\varepsilon}B^{\kappa})\!+2(\partial_{\rho}B^{\rho})^2)+{\rm tdt.},\ \ \nn\\
-\frac{5\hat{h}^2}{ 4}(\partial_{\varepsilon}\hat{B}_{\kappa}\!-\partial_{\kappa}\hat{B}_{\varepsilon})(\partial^{\varepsilon}\hat{B}^{\kappa}\!-\partial^{\kappa}\hat{B}^{\varepsilon})=\frac{5\hat{h}^2}{ 4}(2\hat{B}_{\kappa}(\partial_{\varepsilon}\partial^{\varepsilon}\hat{B}^{\kappa})\!+2(\partial_{\rho}\hat{B}^{\rho})^2)+{\rm tdt.}.\ \ \nn
\eeaa

We could naively try to reduce the complexity  of kinetic terms of other gauge fields  carrying multiple Lorentz indices. We explore if the real massless fields ${B}_{\varepsilon}^{\  (\rho\tau)}$ and ${B}_{\varepsilon}^{\ [\rho\tau]}$ can be cast in terms of real symmetric vector $S_\rho$, $U_\rho$, and  $V_\rho$: 
\beaa
{B}_{\varepsilon}^{\  (\rho\sigma)}&=\delta_{\varepsilon}^{\rho} S^\sigma+\delta_{\varepsilon}^{\sigma} S^\rho-\half g^{\rho\sigma} S_\varepsilon,\hspace{2.5cm}\\
{B}_{\varepsilon}^{\ [\rho\sigma]}&=\tilde{b}(\delta_{\varepsilon}^{\rho} U^\sigma-\delta_{\varepsilon}^{\sigma} U^\rho)+\tilde{a}\epsilon^{\tau\omega\rho\sigma}(g_{\varepsilon\tau}V_\omega-g_{\varepsilon\omega}V_\tau),
\eeaa
using $\epsilon_{0123}=-\epsilon^{0123}=1$. This  would induce kinetic parts given by
\beaa
\!\!-\frac{r^2}{4}(\partial_{\varepsilon}B_{\kappa}^{\, \, (\tau\omega)}\!-\partial_{\kappa}B_{\varepsilon}^{\, \, (\tau\omega)})(\partial^{\varepsilon}B^{\kappa}_{\, \, (\tau\omega)}\!-\partial^{\kappa}B^{\varepsilon}_{\, \, (\tau\omega)})&~\hspace{6cm}&\nn\\
\ \ ={r^2}(4S_{\kappa}(\partial_{\varepsilon}\partial^{\varepsilon}S^{\kappa})\!+2(\partial_{\rho}S^{\rho})^2)+{\rm tdt.},&&\nn\\
\!\!\!\!-\frac{r'^2}{4}(\partial_{\varepsilon}B_{\kappa}^{\, \, [\tau\omega]}\!-\partial_{\kappa}B_{\varepsilon}^{\, \, [\tau\omega]})(\partial^{\varepsilon}B^{\kappa}_{\, \, [\tau\omega]}\!-\partial^{\kappa}B^{\varepsilon}_{\, \, [\tau\omega]})\!&&\nn\\
\hspace{-1cm}\!=\!{r'^2\tilde{b}^2}(2U_{\kappa}(\partial_{\varepsilon}\partial^{\varepsilon}U^{\kappa}\!)\!-\!(\partial_{\rho}U^{\rho})^2)\!+\!{r'^2\tilde{a}^2}&\hspace{-1.2cm}(\!-8V_{\kappa}(\partial_{\varepsilon}\partial^{\varepsilon}V^{\kappa}\!)\!+4(\partial_{\rho}V^{\rho})^2)\!+\!{\rm tdt.}\ \ \ \ &\nn
\eeaa
Such naive trials seem to lead  to no ghosts for the choice $\tilde{b}>0, \tilde{a}=0$, if we maintain the contribution of longitudinal parts $(\delta_{\rho}U^{\rho})^2$ under control, which is done by a proper gauge fixing and Faddeev-Popov or BRST formalism.

\section{Differential Lorentz Extension  Representation}
We explore the representability of the Lorentz Extension (LE) in terms of differential operators in a geometric space, including the 4D space-time coordinates. This will ultimately classify the \textbf{LE as a pseudo--unitary Lie algebra}.

To each symmetry generator \({\cal O}\), we assign a differential operator \(-i\delta_{{\cal O}}\) that, when acting on the fields, represents the same algebra. If \(\lambda^{{\cal O}'}{\cal C}_a^{{\cal O}'}\!(x)\) is a function of the geometric-frame coordinates (such as the representation on the fields, i.e., left Lie-algebra module), then, besides the transformation considered in (\ref{Eqn27})--(\ref{Eqn46}), we introduce a differential operator representation \(-i\delta_{{\cal O}}\) which copes with the dependence on the geometric frame:
\beaa\,[{\cal O},\lambda^{{\cal O}'}{\cal C}_a^{{\cal O}'}]_2&=-\ip\delta_{{\cal O}}(\lambda^{{\cal O}'}{\cal C}_a^{{\cal O}'}),&\nn\\
\ [{\cal O},\lambda^{{\cal O}'}{\cal C}_a^{{\cal O}'}]_1&=-\ip\, C_{\ {\cal O}{\cal O}''}^{{\cal O}'}\lambda^{{\cal O}''}{\cal C}_a^{{\cal O}''},&\label{CommTot}
\eeaa
with \(\lambda^{{\cal O}}=\lambda_{{\cal O}}\) being the coupling associated with \({{\cal O}}\). The indices in the coupling constants do not trigger summations over repeated indices. Since we want a differential representation that commutes with the action of the abstract symmetry generators \({\cal O}\), it turns out that the algebra of differential operators should satisfy
\begin{equation}\label{Algdiff}
[\ip\delta_{{\cal O}} , \ip\delta_{{\cal O}'}] = \ip C^{{\cal O}''}_{{\cal O}{\cal O}'}(\ip\delta_{{\cal O}''} ),\ \text{for}\ [{\cal O},{\cal O}']=\ip C^{{\cal O}''}_{{\cal O}{\cal O}'}{{\cal O}''}.
\end{equation}
There are (at least) two ways of representing the algebra (\ref{Dim0Start})--(\ref{Dim0End}) as differential operators. The diverse generators can either be represented in terms of Hermitian (H) and Anti--Hermitian (AH) operators, or in terms of Hermitian (H) and Adjoint--Pairs (AP) of generators. The nature of generators might change under symmetry breaking. Find above a description of the found differential representations.
\begin{table}[h]
	${\footnotesize
		\bac{c|cccccc}\hline{\stackrel{{\rm Generator}\rightarrow}
			{\downarrow{\rm Diff. Rep.}}} &\ \ \ \ {T}_{\rho}\ \ \ \  &\ \ \  \ \hat{T}_{\rho}\ \ \ \  & \ \ \ \  M_{\rho\tau}\ \  \ \  & \ \ \ \ T_{\rho\tau}\ \ \ \ & \ \ \ \ {T_{[o]}} \ \ \ \ &\ \ \ \ {\hat{T}_{[o]}}\ \ \ \ \\
		\hline
		5D (Brane)&{\rm H} &{\rm H}&{\rm H}&{\rm AH}&{\rm AH}&{\rm AH}\\
		\hline
		10D&{\rm AP} &{\rm AP}&{\rm H}&{\rm H}&{\rm H}&{\rm H}\\
	\hline
	 \eac}$
	\caption{Differential Representations; $\rho,\tau=0,1,2,3.$}\label{DFull}
\end{table}

We first present the Brane or 5D representation in terms of differential operators. In addition to the 4D space--time coordinates \(\chi_{\rho}\) with \(\partial_{\rho}=\partial/\partial \chi^{\rho},\ \rho=0,1,2,3\), we introduce a new variable \(\chi_5\) with \(\partial_5=\partial/\partial \chi^5\), where \(\partial_5 (\chi_5)=\epsilon=g_{55}\). This variable will be associated with the symmetry combining Higgs and gauge fields. We allow for another independent variable \(\chi_4\) with \(\partial_4=\partial/\partial \chi^4\), where \(\partial_4 (\chi_4)=g_{44}\). This is the variable that will lead to an extension of the Poincaré algebra after an optional contraction. Since the coordinate \(\chi_4\) might be removed after the contraction procedure, we do not count this coordinate in the total dimension of the representation.

We represent \(M_{\mu\nu}\) by \(-i(\chi_{\mu}\partial_{\nu}-\chi_{\nu}\partial_{\mu})\), and \(T_{\mu\nu}\) by the Anti-Hermitian operator \(+(\chi_{\mu}\partial_{\nu}+\chi_{\nu}\partial_{\mu}-\frac{1}{2} g_{\mu\nu} \chi^{\rho}\partial_{\rho})\). We adopt \(-i\delta_{T_{\mu}}=+\sqrt{2}i \chi_5\partial_{\mu}\) and \(-i\delta_{\hat{T}_{\nu}}= -\sqrt{2}i \chi_{\nu}\partial_{5}\) to obtain the 5D representation:
 \beaa
 -\ip \delta_{M_{\mu\nu}}&=&\!\!-\ip(\chi_{\mu}\partial_{\nu}-\chi_{\nu}\partial_{\mu}),\
 -\ip\delta_{T_{\mu\nu}}\!=+(\chi_{\mu}\partial_{\nu}\!+\!\chi_{\nu}\partial_{\mu}\!-\!\half g_{\mu\nu} \chi^{\rho}\partial_{\rho}),\ \ \\
 -\ip\delta_{T_{\mu}}&=&\!\!+\sqrt{2}\ip \chi_5\partial_{\mu},\ \ -\ip\delta_{\hat{T}_{\nu}}= -\sqrt{2}\ip \chi_{\nu}\partial_{5},\\
 -\ip\delta_{{T_{[o]}}}&=&\!\!-\frac{5}{6}\epsilon^{-1}\chi_5\partial_5+\frac{1}{6}g_{44}^{-1}\chi_4\partial_4+\frac{1}{6}\chi^{\rho}\partial_{\rho},\\
 -\ip\delta_{{\hat{T}_{[o]}}}&=&-4g_{44}^{-1}\chi_4\partial_4+\chi^{\rho}\partial_{\rho}.
\eeaa
The quantity \(|\chi|^2=\chi^{\rho}\chi_{\rho}+g_{44}\chi^4\chi^4+\epsilon\chi^5\chi^5\), where \(\chi^5=\epsilon^{-1}\chi_5\) and \(\chi^4=g_{44}^{-1}\chi_4\), is not invariant under the Anti--Hermitian differential operators, precluding a simple geometric space interpretation with metric \({\rm diag(1,-1,-1,-1,g_{44},\epsilon)}\) for the LE. However, it is invariant under the subalgebra \(<\{M_{\mu\nu},T_{\rho}+\hat{T}_{\rho}\}>\). Nevertheless, the quantity
\({\cal I}=(\chi^{\mu}\chi_{\mu})^4(\chi^{4}\chi_{4})(\chi^{5}\chi_{5})\), is invariant under \(\delta_{M_{\mu\nu}}, \delta_{{T_{[o]}}}\), and \(\delta_{{\hat{T}_{[o]}}}\). Now, \({\cal I}\) is invariant under \(-i\bar{\xi}^{\mu}\delta_{T_{\mu}}-i\xi^{\mu}\delta_{\hat{T}_{\mu}}\) for \(\xi^{\mu}\) real (or \(+i\bar{\xi}^{\mu}\delta_{T_{\mu}}-i\xi^{\mu}\delta_{\hat{T}_{\mu}}\) for \(\xi^{\mu}\) imaginary) \textbf{and} \(\chi^{5}\chi_{5}=(1/4)\chi^{\mu}\chi_{\mu}\), which looks like an on--shell or perhaps brane condition. For such choices
$-i\bar{\xi}^{\mu}\delta_{T_{\mu}}-i\xi^{\mu}\delta_{\hat{T}_{\mu}}, -i\bar{\kappa}^{\mu}\delta_{T_{\mu}}-i\kappa^{\mu}\delta_{\hat{T}_{\mu}}$
have only nonvanishing contributions proportional to \(-i\delta_{M_{ \mu\nu}}\), which retains invariance. We do not verify invariance of \(|\chi|^2\) under \(\xi^{(\mu\nu)}\delta_{T_{\mu\nu}}\) for arbitrary parameter \(\xi^{(\mu\nu)}\). So, with these tenets, only a subgroup of the original extension seems to be cast with this representation. Therefore, the label of Brane representation since it might be only relevant after some symmetry breaking or some topological reconfiguration into a brane.

The contraction is adjusted to provide a differential representation without involving the \(x^4\) coordinate. We use \(R\rightarrow\infty\), \(\chi_4/R\rightarrow (1/\sqrt{2})\), \(g^{44}\chi_4\partial_4\rightarrow -\epsilon^{-1}\chi_{5}\partial_5-\chi^\rho\partial_{\rho}+3\), and \((1/R)\partial_4\rightarrow 0\).

We naively expect that \(T_{\mu}\) and \(\hat{T}_{\mu}\) (as the Higgs field) build an adjoint pair, as hinted by (\ref{TB}) and (\ref{Eqn27}), but this is not substantiated in the 5D representation. There is a way to restore naive expectations at the expense of complexifying the coordinates. This leads to the 10D representation.

Let \(z_{a}=\chi_{a}+i y_{a}\) and \(\bar{z}_a=\chi_{a}-i y_{a}\) for \(a=0,1,2,3,4,5\). We find a representation with
\beaa \bar{z}^{a}z_{a}&=&g_{\mu\nu}\bar{z}^{\mu}{z}^{\nu}+g_{44}\bar{z}^{4}{z}^{4}+\epsilon\bar{z}^{5}{z}^{5}\nn\\
&=&g_{\mu\nu}{\chi}^{\mu}{\chi}^{\nu}+g_{\mu\nu}{y}^{\mu}{y}^{\nu}+g_{44}{\chi}^{4}{\chi}^{4}+g_{44}{y}^{4}{y}^{4}+\epsilon{\chi}^{5}{\chi}^{5}+\epsilon{y}^{5}{y}^{5}\label{Inv10D}
\eeaa
invariant, giving substance to a relation to a subalgebra of an \({\mathfrak{su}}(s_4,t_4)\) with \(s_4,t_4>0\) and \(s_4+t_4=6\), such as \({\mathfrak{u}}(s_1,t_1)\) with \(s_1,t_1>0\) and \(s_1+t_1=5\). Adopting \(\epsilon\) negative seems to assign a space--like role to the \(\chi^{5}, y^{5}\) coordinates. Let us denote \(\partial_a=\partial/\partial \chi^a\) and \(\hat{\partial}_a=\partial/\partial y^a\). The 10D representation  with such invariance is:
\beaa
-\ip \delta_{M_{\mu\nu}}\!\!\!\!&=&\!\!\!\!-\ip(\chi_{\mu}\partial_{\nu}-\chi_{\nu}\partial_{\mu})-\ip(y_{\mu}\hat{\partial}_{\nu}-y_{\nu}\hat{\partial}_{\mu}),\label{10Dstart}\\
-\ip\delta_{T_{\mu\nu}}\!\!\!\!&=&\!\!\!\!-\ip(\chi_{\mu}\hat{\partial}_{\nu}+\chi_{\nu}\hat{\partial}_{\mu}-\!\half g_{\mu\nu} \chi^{\rho}\hat{\partial}_{\rho})\!+\!\ip(y_{\mu}\partial_{\nu}+y_{\nu}\partial_{\mu}-\!\half g_{\mu\nu} y^{\rho}\partial_{\rho}),\ \ \\
-\ip\delta_{T_{\mu}}\!\!\!\!&=&\!\!\!\!+\frac{\ip}{\sqrt{2}}\left(\! (\chi_5+\!\ip y_5)(\partial_{\mu}-\!\ip\hat{\partial}_{\mu})-(\chi_{\mu}-\!\ip y_{\mu})(\partial_5+\!\ip\hat{\partial}_5)\right)\!,\\
 -\ip\delta_{\hat{T}_{\nu}}\!\!\!\!&=&\!\!\!\!-\frac{\ip}{\sqrt{2}}\left(\!(\chi_{\mu}+\!\ip y_{\mu})(\partial_5-\!\ip\hat{\partial}_5) -(\chi_5-\!\ip y_5)(\partial_{\mu}+\!\ip\hat{\partial}_{\mu})\right)\!,\\
-\ip\delta_{{T_{[o]}}}\!\!\!\!&=&\!\!\!\!-\frac{5\ip}{6}\epsilon^{\!-1}(y_5\partial_5\!-\!\chi_5\hat{\partial}_5)\!+\!\frac{\ip}{6}g_{44}^{-1}(y_4\partial_4\!-\!\chi_4\hat{\partial}_4)\!+\!\frac{\ip}{6}(y^{\rho}\partial_{\rho}\!-\!\chi^{\rho}\hat{\partial}_{\rho}),\ \\
-\ip\delta_{{\hat{T}_{[o]}}}\!\!\!\!&=&\!\!\!\!-4\ip g_{44}^{-1}(y_4\partial_4-\!\chi_4\hat{\partial}_4)+\ip(y^{\rho}\partial_{\rho}-\!\chi^{\rho}\hat{\partial}_{\rho}).\label{10Dend}
\eeaa
The contraction procedure is adjusted to provide a differential representation without the involvement of the \(\chi^4, y^4\) coordinates. Use \(R\rightarrow\infty\), \(\chi_4/R\rightarrow 1/\sqrt{2}\), \(g^{44}(y_4\partial_4-\chi_4\hat{\partial}_4)\rightarrow -\epsilon^{-1}(y_5\partial_5-\chi_5\hat{\partial}_5)-(y^\rho\partial_\rho-\chi^\rho\hat{\partial}_\rho)\), \(y_4/R\rightarrow 0\), \(\partial_4/R\rightarrow 0\), and \(\hat{\partial_4}/R\rightarrow 0\). The invariance of (\ref{Inv10D}) dropping the \(\chi^4\)--coordinate is easily verified. Mechanisms for the transition (before or after contraction) from the 10D to 5D representation will be tackled elsewhere.

We consider a new  base $\tilde{N}_{AB}$ for generators of  10D, with complex derivation $\partial_{z^A}=(1/2)(\partial_{x^A}-\ip\partial_{y^A}),\ 
 \partial_{\bar{z}^A}=(1/2)(\partial_{x^A}+\ip\partial_{y^A})$, and find:
 \beaa
 -\ip\delta_{\tilde{N}_{AB}}&=&-\ip(z_A\partial_{z^B}-\bar{z}_B\partial_{\bar{z}^A}),\ \  A,B=0,1,2,3,4,5.\label{NtildeInic}\\
 \,[\tilde{N}_{AB},\tilde{N}_{CD}]&=&-\ip(g_{AD}\tilde{N}_{CB}-g_{BC}\tilde{N}_{AD}).
 \eeaa
 Lowering and rising of labels via ${\rm diag(1,-1,-1,-1,g_{44},\epsilon)}$. Using (\ref{10Dstart})--(\ref{10Dend}) we relate the old and new basis:
\beaa
M_{\mu\nu}&=&\tilde{N}_{\mu\nu}-\tilde{N}_{\nu\mu},\\
T_{\mu\nu}&=&\ip(\tilde{N}_{\mu\nu}+\tilde{N}_{\nu\mu})-\frac{\ip g_{\mu\nu}}{2}\tilde{N}^{\rho}_{\ \rho},\\
T_{\mu}&=&-\sqrt{2}\tilde{N}_{5\mu},\ \ \hat{T}_{\mu}\ =\ +\sqrt{2}\tilde{N}_{\mu 5},\\
T_{[o]}&=&-\frac{5\ip}{6}\tilde{N}^{4}_{\ 4}+\frac{\ip}{6}\tilde{N}^{5}_{\ 5}+\frac{\ip}{6} \tilde{N}^{\rho}_{\ \rho},\\
\hat{T}_{[o]}&=&-4\ip\tilde{N}^{4}_{\ 4}+\ip \tilde{N}^{\rho}_{\ \rho}.\label{NtildeEnd}
\eeaa
Using this basis we express the quadratic casimir  (\ref{CasimirDim0}) in terms of the new basis:
\beaa
&\frac{1}{4}\,I&= -\tilde{N}^{\mu 5}\tilde{N}_{5\mu}
-\tilde{N}^{5\mu}\tilde{N}_{\mu 5}
-\tilde{N}^{\mu\nu}\tilde{N}_{\nu\mu}
-\tilde{N}^{55}\tilde{N}_{55}
-\tilde{N}^{44}\tilde{N}_{44}\nn\\
&&\ \ +\frac{1}{6}(\tilde{N}^\mu_{\ \mu}+\tilde{N}^4_{\ 4}+\tilde{N}^5_{\ 5})^2.\label{CasimirDim0tildeNLE}
\eeaa
Similar elaborations can be made after contraction.

The set of generators \(\tilde{N}_{AB}\) involved in the \textbf{LE satisfies an \({\mathfrak{u}}(t_1,s_1)\) pseudo--unitary algebra} with \(t_1+s_1=5\), and \(t_1=1\) for \(\epsilon=-1\). The central charge of the LE \(-5{\cal T}=T_{[o]}-\hat{T}_{[o]}\) \(=\ip(4\tilde{N}^4_{\ 4}+\tilde{N}^5_{\ 5}) -(5/6)\ip{\bf D}\), with \({\bf D}\) the complex 6D trace \(\sum_{A\in\{\mu,4,5\}}\tilde{N}^A_{\ A}=\tilde{N}^\mu_{\ \mu}+\tilde{N}^4_{\ 4}+\tilde{N}^5_{\ 5}\) of novel generators in (\ref{NtildeInic}--\ref{NtildeEnd}). The obtained LE quadratic casimir \(I\) in (\ref{CasimirDim0}) and (\ref{CasimirDim0tildeNLE}) lead to conjecture a quartic casimir \(\check{I}\) of the following form after contraction:
\beaa 
&&\frac{1}{2^4}\check{I}= -\!\!\!\!\!\sum_{A,B,C,D\in\{\mu,5\}}\!\!\!\!\!\tilde{N}^{AB}\!\tilde{N}_{BC}\tilde{N}^{CD}\tilde{N}_{DA}.\label{QuarticInv}
\eeaa
In a prospective Poincar\'e extension, the quadratic casimir will be a summation of squares of our adopted generators. Our choices have not been directed to suit just the Lorentz extension. A Poincar\'e extension will have generous possibilities for devising mechanisms affording a transition to a stage corresponding to the standard model prior internal gauge symmetry breaking.

\section{Fields  and  a Locally Invariant Model II}
We address now the use of differential representations for having actual external symmetries transforming also the geometric frame (extended space--time). We notice that the operator \(\partial_{\mu}\) usually inserted in the covariant derivative does not close an algebra with the obtained differential operators \(-\ip\,\delta_{{\cal O}}\). Nevertheless, we observe that the operator \(-\ip\delta_{T_{\mu}}\) in the 5D representation has a close resemblance to the one used to represent translations. It is not ruled out that through some mechanism, it might contribute to the Standard model translations. In this model, we could perhaps abandon the variable \(\chi_4\) altogether since there is already a copy of the Poincaré algebra lurking in the obtained extension, and the LE is not affected by the contraction. We do not know if both Higgs and gravitational interaction could be recovered from such a model, but it is worth pursuing this line of thought for a moment. When introducing further internal symmetries, the \(T_{\mu}\) becomes \(T_{\alpha\mu}\) with \(\alpha\) a fundamental \(SU(2)\) representation label associated with a complex coordinate. So, the variable \(\chi_4\) for getting separately translation generators will be required.

We use either of the differential representations of the previous section to insert the action of the symmetry on the geometric frame with coordinates either \((\chi_0,\chi_1,\chi_2,\chi_3,\chi_4,\chi_5)\), or \((z_0,z_1,z_2,z_3,z_4,z_5, \bar{z}_0,\bar{z}_1,\bar{z}_2,\bar{z}_3,\bar{z}_4,\bar{z}_5)\), and the fields are functions thereof. We are forced to depart from the standard gauge field formalism and develop a novel formalism.

We propose finally, as a covariant derivative, a naive extrapolation of the gauge formalism. In gauge theories, the covariant derivatives are deformations of the differential representation of the (mutually commuting) Poincaré translation operators. So, we construct a gauge theory over the Lie group manifold itself, in the expectation that further contraction or symmetry-breaking procedures will land in the 3+1-dimensional Minkowski space. The fields and field strengths have now both a naive energy dimension 0 and will resemble a representation metric array and structure constants, respectively:
\beaa
{D}_{{\cal O}}\!\!\!\!&:=& \ip(-\ip\delta_{{\cal O}})\otimes Id+\sum_{\tilde{\cal O}}\ip\lambda^{\tilde{\cal O}}{\cal A}_{{\cal O}}^{\ \tilde{\cal O}}\otimes {\tilde{\cal O}},\\
\sum_{\tilde{\cal O}}\ip\lambda^{\tilde{\cal O}}{\bf F}_{\ {\cal O}{\cal O}'}^{\tilde{\cal O}}\otimes {\tilde{\cal O}}\!\!&:=& \,[{D}_{{\cal O}},{D}_{{\cal O}'}]+\ip {D}_{[{\cal O},{\cal O}']},\label{NewFS}
\eeaa
\beaa
\lambda^{\hat{\cal O}}{\bf F}_{\ {\cal O}{\cal O}'}^{\hat{\cal O}}:=\delta_{{\cal O}}(\lambda^{\hat{\cal O}}{\cal A}_{{\cal O}'}^{\ \hat{\cal O}})-\delta_{{\cal O}'}(\lambda^{\hat{\cal O}}{\cal A}_{{\cal O}}^{\ \hat{\cal O}}) +\lambda^{\hat{\cal O}} {\bf{\cal F}}_{\ {\cal O}{\cal O}'}^{\hat{\cal O}}+\ip(\ip C_{\ {\cal O}{\cal O}'}^{{\cal O}''})(\lambda^{\hat{\cal O}}{\cal A}_{{\cal O}''}^{\ \hat{\cal O}}),\label{NewFS1}\\
\lambda^{\hat{\cal O}}{\bf{\cal F}}_{\ {\cal O}{\cal O}'}^{\hat{\cal O}}:={-\ip}(\lambda^{\tilde{\cal O}}{\cal A}_{{\cal O}}^{\ \tilde{\cal O}})[\tilde{{\cal O}},\lambda^{\hat{\cal O}}{\cal A}_{{\cal O}'}^{\ \hat{\cal O}}(x)]_1={\ip}(\lambda^{\tilde{\cal O}}{\cal A}_{{\cal O}}^{\ \tilde{\cal O}})\ip C_{\ \tilde{\cal O}{\cal O}''}^{\hat{\cal O}}
(\lambda^{{\cal O}''}{\cal A}_{{\cal O}'}^{\ {\cal O}''}).\ \ 
\eeaa
Notice that the curvature in (\ref{NewFS}) involves, besides bilinear terms (associated with the commutation of covariant derivatives), also a term linear in covariant derivatives which might give rise to quadratic (mass--) terms in an invariant Lagrangian quadratic in field strengths. The calligraphic--noted term $\lambda^{\hat{\cal O}}{\bf{\cal F}}_{\ {\cal O}{\cal O}'}^{\hat{\cal O}}$ is again quadratic in the fields and coincides with the expressions in (\ref{QuadraticPartInic}--\ref{QuadraticPartEnd}), changing the indices $\varepsilon, \kappa$ by generator labels ${\cal O}, {\cal O}'$.
Indices ${\cal O}, {\cal O}',$ etc. run over the generators of the Lorentz extension, and the differential operators $-\ip\delta_{{\cal O}}$ are those of the previous section. Later on, one can consider the restriction to covariant derivatives ${D}_{{\cal O}}$ for ${\cal O}$ generators of an invariant subalgebra. We maintain the Einstein summation convention over upper and lower repeated indices (except for those in coupling constants). The $\lambda^{\tilde{\cal O}}$, ${\cal A}_{{\cal O}}^{\ \tilde{\cal O}}(x)$, $ {\bf F}_{\ {\cal O}{\cal O}'}^{\tilde{\cal O}}(x)$ are respectively the coupling constant, field, and field strength, and $(x)$ stands for coordinates in the adopted differential representation. Both the field and field strength are left--modules over the extended Lorentz Lie algebra, transforming only with respect to the upper index:
\beaa \ [\tilde{{\cal O}},\lambda^{\hat{\cal O}}{\cal A}_{{\cal O}}^{\ \hat{\cal O}}(x)]_1&=-\ip\, C_{\ \tilde{{\cal O}}{\cal O}''}^{\hat{{\cal O}}}\lambda^{{\cal O}''}{\cal A}_{{\cal O}}^{\ {\cal O}''}(x),&\nn\\
\ [\tilde{{\cal O}},\lambda^{\hat{\cal O}} {\bf F}_{\ {\cal O}{\cal O}'}^{\hat{\cal O}}(x)]_1&=-\ip\, C_{\ \tilde{{\cal O}}{\cal O}''}^{\hat{\cal O}}\lambda^{{\cal O}''} {\bf F}_{\ {\cal O}{\cal O}'}^{{\cal O}''}(x).&\label{LeftModNew}
\eeaa
Using the  matrix notation ($Id$ being the identity matrix),
\beaa
\,[D_{\cal O}]_{\hat{\cal O}'}^{\ \ \hat{\cal O}}=\delta_{\cal O}[Id]_{\hat{\cal O}'}^{\ \ \hat{\cal O}}+\ip (\lambda^{\tilde{\cal O}}{\cal A}_{\cal O}^{\ \tilde{\cal O}})[R_{\tilde{\cal O}}]_{\hat{\cal O}'}^{\ \ \hat{\cal O}},\\
{\rm with}\ \ [R_{\cal O}]_{\hat{\cal O}'}^{\ \ \hat{\cal O}}=-\ip C_{\ {\cal O}\hat{\cal O}'}^{\hat{\cal O}},
\eeaa
we can recover the results in (\ref{NewFS}--\ref{NewFS1}). $R_{\cal O}$ is the well known adjoint representation.

We generalize the Ansatz for first order gauge transformations with $(x)$--dependent parameters $\varsigma(x),\hat{\varsigma}(x)$, $\hat{\varsigma}^{\nu}(x),\varsigma^{\nu}(x)$, $\varsigma^{[\tau\kappa]}(x),\varsigma^{(\tau\kappa)}(x)$ associated respectively to generators  ${T_{[o]}}, \hat{T}_{[o]}, T_{\nu}$, $\hat{T}_{\nu}, M_{\tau\kappa}$, $T_{\tau\kappa}$:
\beaa 
\!\!(\lambda^{\hat{{\cal O}}}{{\cal A}_{{\cal O}}^{\ \hat{\cal O}}})\!\mapsto\!(\lambda^{\hat{{\cal O}}}{{\cal A}_{{\cal O}}^{\ \hat{\cal O}}})'\!\!=\!(\lambda^{\hat{{\cal O}}}{{\cal A}_{{\cal O}}^{\ \hat{\cal O}}})\!+\!\sum_{\tilde{{\cal O}}}\!\ip{\varsigma}^{\tilde{{\cal O}}}(x) [\tilde{{\cal O}},\lambda^{\hat{\cal O}}{\cal A}_{{\cal O}}^{\ \hat{{\cal O}}}]_1\!+\!%\nn\\
\ip(-\ip\delta_{{\cal O}}\varsigma^{\hat{{\cal O}}}(x)).\ \label{NewFOTField}
\eeaa
From this, the field strengths end up transforming homogeneously (as a curvature), as they would with global (rigid) transformations:
\beaa
(\lambda^{\hat{\cal O}} {\bf F}_{\ {\cal O}{\cal O}'}^{\hat{\cal O}})\mapsto (\lambda^{\hat{\cal O}} {\bf F}_{\ {\cal O}{\cal O}'}^{\hat{\cal O}})'= (\lambda^{\hat{\cal O}} {\bf F}_{\ {\cal O}{\cal O}'}^{\hat{\cal O}})+\sum_{\tilde{{\cal O}}}\ip{\varsigma}^{\tilde{{\cal O}}}(x) [\tilde{{\cal O}},\lambda^{\hat{\cal O}} {\bf F}_{\ {\cal O}{\cal O}'}^{\hat{\cal O}}]_1.\ \ \label{NewFOTFieldSt}
\eeaa
Using the assignment of fields in previous sections, and dropping tensor--products to alleviate notation:
\begin{flalign}
&{ D}_{{\cal O}}=\!\ip(-\ip\delta_{{\cal O}})+\ip \ell (\hat{\Phi}_{{\cal O}}^{\, \, \tau}{T_{\tau}}+\Phi_{{\cal O}}^{\, \, \tau} \hat{T}_{\tau})
+\ip h B_{{\cal O}}T_{[o]}+\ip \hat{h} \hat{B}_{{\cal O}}\hat{T}_{[o]}\hspace{1.5cm}&\nn\\
&\hspace{1.5cm}+\ip r' B^{\, \, [\rho\sigma]}_{{\cal O}}M_{\rho\sigma}+\ip r B^{\, \, (\rho\sigma)}_{{\cal O}}T_{\rho\sigma},&\\
&\,\!\![{ D}_{{\cal O}},{ D}_{{\cal O}'}]+\ip {D}_{[{\cal O},{\cal O}']}\!=\!\ip h\,{ F}_{{\cal O}{\cal O}'}\,T_{[o]}\!+\! \ip \hat{h}\,\hat{{ F}}_{{\cal O}{\cal O}'}\,\hat{T}_{[o]}+
\ip  \ell\, \hat{{ F}}_{\ {\cal O}{\cal O}'}^{\rho}\, T_{\rho} +\ip \ell\,{ F}_{\ {\cal O}{\cal O}'}^{\rho}\,  \hat{T}_{\rho}\hspace{0.8cm}& \nn\\
&\hspace{1.5cm}+ \ip  r'\,{ G}_{\ \ \ {\cal O}{\cal O}'}^{[\tau\kappa]} M_{\tau\kappa}+\ip r { H}_{\ \ \ {\cal O}{\cal O}'}^{(\tau\kappa)}  T_{\tau\kappa}.\!\!&
\end{flalign}
Additional contributions appear from the previously obtained for generic tilded field strengths appear -- compare with (\ref{FS}) -- due to the non--commutativity of differential operators in (\ref{NewFS}). These new terms generate quadratic mass terms.

The action in (\ref{LeftModNew}) on the fields and field strengths is given by the expressions in (\ref{Eqn27}--\ref{Eqn46}) by replacing the index $\kappa$ by ${\cal O}$; and similar expressions changing fields by field strengths. The first-order transformations of the fields ${\cal A}_{{\cal O}}^{\ \hat{\cal O}}$ in (\ref{NewFOTField}) are obtained from (\ref{FOTInic}--\ref{FOTEnd}) by replacing the index $\kappa$ by ${\cal O}$, and changing $\partial_\kappa$ by $\delta_{{\cal O}}$. The first-order transformations of the field strengths ${\bf F}_{\ {\cal O}{\cal O}'}^{\hat{\cal O}}$ in (\ref{NewFOTFieldSt}) are obtained similarly by changing fields by field strengths and dropping the inhomogeneous terms (those with $\delta_{{\cal O}}$).

Additionally, the wider spectrum of covariant derivatives allows for a different reading of the actions of generators on fields: The transformations (\ref{Eqn27}-\ref{Eqn28}) and (\ref{Eqn32}) when the label $b$ is a scalar operator ${\cal O}$ represent rather a mixing between the scalar fields $hB_{\cal O}$,  $h\hat{B}_{\cal O}$ and the vector-like fields $\ell\Phi_{\cal O}^{\tau}$, $\ell\bar{\Phi}_{\cal O}^{\tau}$, which embody the aimed transformations in (\ref{THbar}-\ref{TA}) for the case with just $U(1)$--symmetry.

As we see, derivative-laden terms appear in each $\delta_{{\cal O}}$ in the field strength. This is not unexpected from the classical point of view; for instance, rotational contributions to the kinetic term are expected. The field strengths ${\bf F}_{\ {\cal O}{\cal O}'}^{\hat{\cal O}}$ have an upper index $\hat{\cal O}$ which transforms as in (\ref{NewFOTField}) and whose basis is the interaction basis, hence the coupling constants. The lower indices ${\cal O}{\cal O}'$ are those associated with the covariant derivatives and are not in the interaction basis. To construct an invariant, we contract the upper indices $\hat{\cal O}, \hat{\cal O}'$ using, besides the Minkowski metric (for lowering or rising indices), also the coefficients of $M^{-1}$ in (\ref{CasimirDim0Int}) depending on the index type, completing a matrix ${\bf M}^{-1}_{\hat{\cal O}\hat{\cal O}'}$. For contracting lower indices ${\cal O},{\cal O}''$, besides the Minkowski metric for lowering or rising Lorentz indices, also use coefficients of $M$ in (\ref{CasimirDim0}), completing a matrix ${\bf M}^{{\cal O}{\cal O}''}$. The dual or adjoint basis $v^{\cal O}={\bf M}^{{\cal O}{\cal O}''}{\cal O}''$ transforms $[\hat{\cal O},v^{\cal O}]=-\ip C_{\ \hat{\cal O}\tilde{\cal O}}^{{\cal O}}v^{\tilde{\cal O}}$ with the adjoint representation as does the field strength. From the invariance of $v^{\cal O}{\bf M}^{-1}_{{\cal O}{\cal O}'}v^{{\cal O}'}$, it follows $C^{\tilde{\cal O}}_{\ \hat{\cal O}{\cal O}}{\bf M}^{-1}_{\tilde{\cal O}{\cal O}'} =-{\bf M}^{-1}_{{\cal O}\tilde{\cal O}}C^{\tilde{\cal O}}_{\ \hat{\cal O}{\cal O}'}$, making the array $C_{{\cal O}\hat{\cal O}{\cal O}'}$ totally antisymmetric. Hence, using the dual counterpart formalism to build a globally invariant Lagrangian, we obtain:
\beaa
&&\!\!\!\!\!\!4{L}_o\!=\!\ip\!\lambda^{\hat{\cal O}}{\bf F}_{\ {\cal O}{\cal O}'}^{\hat{\cal O}}{\bf M}^{-1}_{\hat{\cal O}\hat{\cal O}'}{\bf M}^{{\cal O}{\cal O}''}{\bf M}^{{\cal O}'{\cal O}'''}\!\ip\!\lambda^{\hat{\cal O}'}{\bf F}_{\ {\cal O}''{\cal O}'''}^{\hat{\cal O}'}\!=\!-(\lambda^{\hat{\cal O}})^2{\bf F}_{\ {\cal O}{\cal O}'}^{\hat{\cal O}}{\bf F}^{\ {\cal O}{\cal O}'}_{\hat{\cal O}}=\!\!\!\!\nn\\
&&\!\!\!\!\!\!\ip[\ell{{\bar{F}}}_{\ {\cal O}{\cal O}'}^{ \kappa}g_{\kappa\kappa'}\ \ell{{F}}_{\ {\cal O}{\cal O}'}^{\kappa}g_{\kappa\kappa'}\ r'{{G}}_{\ \ \ {\cal O}{\cal O}'}^{[\rho\tau]}\!g_{\rho\rho'}g_{\tau\tau'}\ r{{H}}_{\ \ \ {\cal O}{\cal O}'}^{(\rho\tau)}\ h{{F}}_{{\cal O}{\cal O}'}\!g_{\rho\rho'}g_{\tau\tau'}\ \hat{h}{{\hat{F}}}_{{\cal O}{\cal O}'}]\cdot \hspace{1cm}\ \nn\\
&&M^{-1}\cdot\ip[\ell{{F}}_{\ {\cal O}''{\cal O}'''}^{ \kappa'}\ \ell{{\bar{F}}}_{\ {\cal O}''{\cal O}'''}^{ \kappa'}\ r'{{G}}_{\ \ \ {\cal O}''{\cal O}'''}^{ [\rho'\tau']}\ r{{H}}_{\ \ \ {\cal O}''{\cal O}'''}^{ (\rho'\tau')}\ h{{F}}_{{\cal O}''{\cal O}'''}\ \hat{h}{{\hat{F}}}_{{\cal O}''{\cal O}'''}]^{tr}\!\!\!\!\nn \\
&&{\bf M}^{{\cal O}{\cal O}''}\!{\bf M}^{{\cal O}'{\cal O}'''},\!\!\nn
\eeaa
\beaa
&&4L_o=\left( -\frac{\epsilon\ell^2}{2}\{ {{\bar{F}}}^{\ {\cal O}{\cal O}'}_{ \kappa},{{F}}_{\ {\cal O}{\cal O}'}^{\kappa}\}-{r'^2}{{G}}^{\ \ \ {\cal O}{\cal O}'}_{[\rho\tau]}{{G}}_{\ \ \ {\cal O}{\cal O}'}^{[\rho\tau]}- 
{r^2}{{H}}^{{\cal O}{\cal O}'} _{\ \ (\rho\tau)}{{H}}_{{\cal O}{\cal O}'}^{\ \ (\rho\tau)}\right.\nn \\
&&\hspace{2cm}\left.- \frac{5h^2}{24}{{F}}^{{\cal O}{\cal O}'}{{F}}_{{\cal O}{\cal O}'}-   {5\hat{h}^2}{{\hat{F}}}^{{\cal O}{\cal O}'}{\hat{F}}_{{\cal O}{\cal O}'}\right).\label{InvLagr}
\eeaa
 The local invariance of the action with Lagrangian $ {L}_o$ follows from the fact that  field strengths transform homogeneously, and the quadratic casimir structure of the action does the rest. So, we have for instance $[\ell\,T_{\mu}, \int\cdots\int {L}_o dx^n]=0$.

\section{Equations of Motions \& Gauge Invariance}
In this section, we provide the field theoretical considerations leading to the equations of motion and verify again the invariance of the Action with Lagrangian quadratic in the field strengths. We determine as well the conserved currents, as well as covariantly conserved currents. Such preliminary developments are necessary to consider the model as a suitable candidate for a quantum field theory. Observe that we succeed in creating a model with Local invariance under external symmetries involving Lorentz transformations.

First, we should notice that our group does not have generators for actual plain translations in the adopted extended space--time, nor any device has been set up yet to gain them after, say, a contraction mechanism or some analogous procedure. So, we are at a stage where plain space--time translations are not allowed.

Let $F(x)$, with $(x)$ standing for the coordinates involved in the adopted differential representation $\{-\ip\delta_{{\cal O}}\,|\, {{\cal O}}$ operator of the extended Lorentz symmetry$\}$, be a differentiable function of its arguments, and that vanishes or is exponentially decaying on each of the limits of the integration of the action $\int\cdots\int {L}_o dx^n$. Let $\tilde{\varsigma}(x)$ be an arbitrary  continuously differentiable function of the geometric--frame coordinates. For the obtained differential representations, we have:
\beaa
\int\cdots\int F(x)\,(-\ip\delta_{{\cal O}"}\tilde{\varsigma}(x))\, d\chi^n=
-\int\cdots\int(-\ip\delta_{{\cal O}"} F(x))\tilde{\varsigma}(x)\, d\chi^n\nn\\
+({\rm vanishing\  boundary\  contributions}).\label{Adjoint}
\eeaa 
This property holds for the differential representations in the previous section independently of the typology of the $-\ip\delta_{{\cal O}"}$, be it hermitian, anti-hermitian, or a member of an adjoint pair.

Now, ${{ L}_o}(\lambda^{\hat{\cal O}}{{\cal A}}_{{\cal O}}^{\ \hat{\cal O}}(x), -\ip\delta_{\tilde{\cal O}}\lambda^{\hat{\cal O}}{{\cal A}}_{{\cal O}}^{\ \hat{\cal O}}(x))$ is a differentiable function of its arguments, and it vanishes for vanishing arguments. We assume that both the fields $\lambda^{\hat{\cal O}}{{\cal A}}_{{\cal O}}^{\ \hat{\cal O}}(x)$ (accompanied with its coupling constant), and $-\ip\delta_{\tilde{\cal O}}\lambda^{\hat{\cal O}}{{\cal A}}_{{\cal O}}^{\ \hat{\cal O}}(x)$ vanish or are exponentially decaying on each of the integration limits of the action $A_o=\int\cdots\int {L}_o d\chi^n$. We use (\ref{Adjoint}) and the variational analysis notation.
\bee{VarLo}
\frac{\delta A_o}{\delta(\lambda^{\hat{\cal O}}{{\cal A}}_{{\cal O}}^{\ \hat{\cal O}}(x))}=\frac{\partial{{ L}_o} }{\partial(\lambda^{\hat{\cal O}}{{\cal A}}_{{\cal O}}^{\ \hat{\cal O}}(x))} -\left(-\ip\delta_{\tilde{\cal O}}\,\frac{\partial{{ L}_o}}{\partial( -\ip\delta_{\tilde{\cal O}}\lambda^{\hat{\cal O}}{{\cal A}}_{{\cal O}}^{\ \hat{\cal O}}(x))}\right).\ee
We have  ${ L}_o=\frac{1}{4}(\ip\lambda^{\hat{\cal O}}{\bf F}_{\ {\cal O}{\cal O}'}^{\hat{\cal O}})(\ip\lambda_{\hat{\cal O}}{\bf F}^{\ {\cal O}{\cal O}'}_{\hat{\cal O}})$. 
The condition of a vanishing variation of $A_o$ under arbitrary field variations (consistent with their conditions at the integration limits) becomes (in the first line we use (\ref{VarLo})):
\beaa
&&\delta A_o=\int\cdots\int \left(\delta(\lambda^{\hat{\cal O}}{{\cal A}}_{{\cal O}}^{\ \hat{\cal O}}(x))\frac{\delta A_o}{\delta(\lambda^{\hat{\cal O}}{{\cal A}}_{{\cal O}}^{\ \hat{\cal O}}(x))}\right)d\chi^n+({\rm v.\  b.\  c.})\!\!\!\nn\\
&&\!=\!\!\int\cdots\int \delta(\lambda^{\hat{\cal O}}{{\cal A}}_{{\cal O}}^{\ \hat{\cal O}}(x))\left(\frac{\partial(\lambda^{\hat{\cal O}}{\bf F}_{\ {\cal O}{\cal O}'}^{\hat{\cal O}})}{\partial(\lambda^{\hat{\cal O}}{{\cal A}}_{{\cal O}}^{\ \hat{\cal O}}(x))}\frac{\partial{{ L}_o}\ \ \  }{\partial(\lambda^{\hat{\cal O}}{\bf F}_{\ {\cal O}{\cal O}'}^{ \hat{\cal O}})}\right.\nn\\
&& \left.-(-\ip\delta_{\tilde{\cal O}})\left(\,\frac{\partial(\lambda^{\hat{\cal O}}{\bf F}_{\ {\cal O}{\cal O}'}^{\hat{\cal O}})\ \ }{\partial( -\ip\delta_{\tilde{\cal O}}\lambda^{\hat{\cal O}}{{\cal A}}_{{\cal O}}^{\ \hat{\cal O}}(x))}\frac{\partial{{ L}_o}\ \ \  }{\partial(\lambda^{\hat{\cal O}}{\bf F}_{\ {\cal O}{\cal O}'}^{\hat{\cal O}})}\right)\right)\!d\chi^n+({\rm v.\  b.\  c.})=0,\!\!\!\nn 
\eeaa
where v. b. c. stands for vanishing boundary contributions. 
The {\bf equation of motion} guarantees the invariance of the action for generic field variations:
\bee{EqMot}
\,[D_{{\cal O}'}]_{\hat{\cal O}'}^{\ \ \hat{\cal O}}\left({\lambda_{\hat{\cal O}}}{\bf F}_{\hat{\cal O}}^{\ {\cal O}{\cal O}'}\right)-\ip(\ip C_{\ \bar{\cal O}{\cal O}'}^{{\cal O}})\left({\lambda_{\hat{\cal O}'}}{\bf F}_{\hat{\cal O}'}^{\ \bar{\cal O}{\cal O}'}\right)=0.
\ee
In terms of the field strength ${\bf F}_{\ {\cal O}{\cal O}'}^{\hat{\cal O}}$  the equation of motion becomes
\beaa
\,[D^{{\cal O}'}]^{\hat{\cal O}'}_{\ \ \hat{\cal O}}\left({\lambda^{\hat{\cal O}}}{\bf F}^{\hat{\cal O}}_{\ {\cal O}{\cal O}'}\right)-\ip(\ip C^{\ \bar{\cal O}{\cal O}'}_{{\cal O}})\left({\lambda^{\hat{\cal O}'}}{\bf F}^{\hat{\cal O}'}_{\ \bar{\cal O}{\cal O}'}\right)=0,\hspace{2cm}&&\\
{\rm where}\ [D^{{\cal O}'}]^{\hat{\cal O}'}_{\ \ \hat{\cal O}}=\delta^{{\cal O}'}[Id]^{\hat{\cal O}'}_{\ \ \hat{\cal O}}+\ip (\lambda_{\tilde{\cal O}}{\cal A}^{{\cal O}'}_{\ \tilde{\cal O}})[R^{\tilde{\cal O}}]^{\hat{\cal O}'}_{\ \ \hat{\cal O}},\ 
 [R^{\tilde{\cal O}'}]^{\hat{\cal O}'}_{\ \ \hat{\cal O}}=-\ip C^{\ \tilde{\cal O}\hat{\cal O}'}_{\hat{\cal O}},\ &&
\eeaa
lowering and rising of indices are done using ${\bf M}^{-1}_{\hat{\cal O}\hat{\cal O}'}$ and  ${\bf M}^{{\cal O}{\cal O}''}$ respectively. Notice, $C^{\ \tilde{\cal O}\hat{\cal O}'}_{\hat{\cal O}}$ is the structure constant of the dual basis algebra, $v^{\cal O}={\bf M}^{{\cal O}{\cal O}''}{\cal O}''$.

From the equation of motion (\ref{EqMot})  it is easy to propose a  conserved current, where the last equality in (\ref{Current}) uses the equation of motion:
\beaa
&&\!\!{\cal J}_{\hat{\cal O} }^{\ \bar{\cal O} }=\delta_{{\cal O}'}(\lambda_{\hat{\cal O}}{\bf F}_{\hat{\cal O}}^{\ \bar{\cal O}{\cal O}'})-\frac{\ip}{2}(\ip C_{\ {\cal O}{\cal O}'}^{\bar{\cal O}})(\lambda_{\hat{\cal O}}{\bf F}_{\hat{\cal O}}^{\ {\cal O}{\cal O}'})\nn\\
&&\ \ =-\ip(\lambda^{\hat{\cal O}'}{{\cal A}}_{{\cal O}'}^{\ \hat{\cal O}'})(-\ip C_{\ \hat{\cal O}'\hat{\cal O}}^{\tilde{\cal O}})(\lambda_{\tilde{\cal O}}{\bf F}_{\tilde{\cal O}}^{\ \bar{\cal O}{\cal O}'})+\frac{\ip}{2}(\ip C_{\ {\cal O}{\cal O}'}^{\bar{\cal O}})(\lambda_{\hat{\cal O}}{\bf F}_{\hat{\cal O}}^{\ {\cal O}{\cal O}'}),\ \ \ \  \label{Current}\\
&&\delta_{\bar{\cal O}}{\cal J}_{\hat{\cal O} }^{\ \bar{\cal O}}=0.
\eeaa
We can also define a covariantly conserved current, where the last equality in (\ref{CovCurrent}) uses the equation of motion:
\beaa
\!\!{ J}_{\hat{\cal O} }^{\ \bar{\cal O} }=[D_{{\cal O}'}]_{\hat{\cal O}}^{\ \tilde{\cal O}}(\lambda_{\tilde{\cal O}}{\bf F}_{\tilde{\cal O}}^{\ \bar{\cal O}{\cal O}'})\!-\!\frac{\ip}{2}(\ip C_{\ {\cal O}{\cal O}'}^{\bar{\cal O}})(\lambda_{\hat{\cal O}}{\bf F}_{\hat{\cal O}}^{\ {\cal O}{\cal O}'}) =\frac{\ip}{2}(\ip C_{\ {\cal O}{\cal O}'}^{\bar{\cal O}})\lambda_{\hat{\cal O}}{\bf F}_{\hat{\cal O}}^{\ {\cal O}{\cal O}'}\!\!,\label{CovCurrent}\\
\,[D_{\bar{\cal O}}]_{\hat{\cal O}}^{\ \tilde{\cal O}}{ J}_{\tilde{\cal O} }^{\ \bar{\cal O}}=0.\hspace{8cm}
\eeaa
Now, the dynamic of our model thus far is governed by the underlying group and algebra. We know which is the transformation of fields and their corresponding field strengths under local group actions. We consider the  action of differential representations on the extended space--time, constraining them to produce a smooth bijection, so that the change of variables theorem apply:
\beaa
\chi^a&\mapsto& \chi^a-\ip\vartheta \varsigma^{\tilde{\cal O}}(x)(-\ip\delta_{\tilde{\cal O}} \chi^a),\\
d\chi^a&\mapsto&d\chi^a+\partial_b(-\ip\vartheta \varsigma^{\tilde{\cal O}}(x)(-\ip\delta_{\tilde{\cal O}}\chi^a))d\chi^b,\nn\\
d\chi^n&\mapsto&d\chi^n+\partial_a(-\ip \vartheta\varsigma^{\tilde{\cal O}}(x)(-\ip\delta_{\tilde{\cal O}}\chi^a))d\chi^n.\label{VarMeasure}
\eeaa
Notice that only diagonal contributions contribute in first order to the full volume measure $d\chi^n$.
We can turn on and off the action of the differential representation taking $\vartheta=1$ or $\vartheta=0$. We examine dynamical conditions that assure the  first order invariance of the action under local gauge transformations (for $\vartheta=0$), or a combination of both the differential representation and local gauge transformations
 (for $\vartheta=1$).
 In  (\ref{VarField}),  the field $\lambda^{\hat{\cal O}}{{\cal A}}_{{\cal O}}^{\ \hat{\cal O}}(x)$  transforms as $\lambda^{\hat{\cal O}}{{\cal A}}_{{\cal O}}^{\ \hat{\cal O}}(x)\mapsto \lambda^{\hat{\cal O}}{{\cal A}}_{{\cal O}}^{\ \hat{\cal O}}(x)+ \delta(\lambda^{\hat{\cal O}}{{\cal A}}_{{\cal O}}^{\ \hat{\cal O}}(x))$   with  
\beaa
\delta(\lambda^{\hat{\cal O}}{{\cal A}}_{{\cal O}}^{\ \hat{\cal O}}(x))=\sum_{\tilde{\cal O}}\ip\vartheta{{\varsigma}^{\tilde{\cal O}}(x)} (\!-(-\ip\delta_{\tilde{\cal O}}(\lambda^{\hat{\cal O}}{{\cal A}}_{{\cal O}}^{\ \hat{\cal O}}(x))))\hspace{2cm}\nn\\
\hspace{2cm}+\sum_{\tilde{\cal O}}\ip{{\varsigma}^{\tilde{\cal O}}(x)}[\tilde{\cal O},\lambda^{\hat{\cal O}}{{\cal A}}_{{\cal O}}^{\ \hat{\cal O}}(x)]_1 +\ip(-\ip\delta_{{\cal O}}\varsigma^{\hat{{\cal O}}}(x)).&&\label{VarField}
\eeaa
 
 Using (\ref{Adjoint}),
the condition of a vanishing variation of $A_o$ under local gauge transformations (\ref{VarField}) and (\ref{VarMeasure}) becomes:
\beaa
\int\cdots\int \left(\delta(\lambda^{\hat{\cal O}}{{\cal A}}_{{\cal O}}^{\ \hat{\cal O}}(x))\frac{\delta A_o}{\delta(\lambda^{\hat{\cal O}}{{\cal A}}_{{\cal O}}^{\ \hat{\cal O}}(x))}+L_o \partial_a(-\ip\vartheta \varsigma^{\tilde{\cal O}}(x)(-\ip\delta_{\tilde{\cal O}}x^a))\right)d\chi^n=\!\!\!\nn&\\
\int\cdots\int \left(\delta(\lambda^{\hat{\cal O}}{{\cal A}}_{{\cal O}}^{\ \hat{\cal O}}(x))\frac{\delta A_o}{\delta(\lambda^{\hat{\cal O}}{{\cal A}}_{{\cal O}}^{\ \hat{\cal O}}(x))}-\partial_a(L_o) (-\ip\vartheta \varsigma^{\tilde{\cal O}}(x)(-\ip\delta_{\tilde{\cal O}}x^a))\right)d\chi^n\!\!\!\nn&\\ 
+({\rm vanishing\  boundary\  contributions})=&\nn\\
\int\cdots\int \left(\delta(\lambda^{\hat{\cal O}}{{\cal A}}_{{\cal O}}^{\ \hat{\cal O}}(x))\frac{\delta A_o}{\delta(\lambda^{\hat{\cal O}}{{\cal A}}_{{\cal O}}^{\ \hat{\cal O}}(x))}-(-\ip\vartheta \varsigma^{\tilde{\cal O}}(x))(-\ip\delta_{\tilde{\cal O}}(L_o)) \right)d\chi^n\ \ \nn&\\ 
+({\rm vanishing\  boundary\  contributions})=0.\ \ &
\eeaa
Now, we have  ${ L}_o=(\ip\!\lambda^{\hat{\cal O}})^2{\bf F}_{\ {\cal O}{\cal O}'}^{\hat{\cal O}}{\bf F}^{\ {\cal O}{\cal O}'}_{\hat{\cal O}}$. The variation of field strengths\\ $(\lambda^{\hat{\cal O}}{\bf F}_{\ {\cal O}{\cal O}'}^{ \hat{\cal O}})$ $\mapsto (\lambda^{\hat{\cal O}}{\bf F}_{\ {\cal O}{\cal O}'}^{ \hat{\cal O}})+\delta(\lambda^{\hat{\cal O}}{\bf F}_{\ {\cal O}{\cal O}'}^{ \hat{\cal O}})$  is: 
\beaa
\!\!\!\delta(\lambda^{\hat{\cal O}}{\bf F}_{\ {\cal O}{\cal O}'}^{ \hat{\cal O}})\!\!&\!\!\!=\!
\sum_{\tilde{\cal O}}\left(\!-\ip\vartheta{{\varsigma}^{\tilde{\cal O}}\!(x)} (-\ip\delta_{\tilde{\cal O}}(\lambda^{\hat{\cal O}}{\bf F}_{\ {\cal O}{\cal O}'}^{\hat{\cal O}})\!\!\right)\!\nn\\
&+\!\sum_{\tilde{\cal O}}\!\ip{{\varsigma}^{\tilde{\cal O}}\!(x)}[\tilde{\cal O},{\lambda^{\hat{\cal O}}}{\bf F}_{\ {\cal O}{\cal O}'}^{\hat{\cal O}}]_1.\ \ \ \label{VarFS}
\eeaa
Hence, the condition for vanishing variation of the action $A_o$ becomes
\beaa
&&\int\!\cdots\!\int\! \left(\delta(\lambda^{\hat{\cal O}}{\bf F}_{\ {\cal O}{\cal O}'}^{ \hat{\cal O}})\frac{\delta A_o}{\delta(\lambda^{\hat{\cal O}}{\bf F}_{\ {\cal O}{\cal O}'}^{ \hat{\cal O}})}-(-\ip\vartheta \varsigma^{\tilde{\cal O}}(x))(-\ip\delta_{\tilde{\cal O}}(L_o) \right)d\chi^n\nn\\ 
&&=\int\!\cdots\!\int\!\left(\left(\!-\ip\vartheta{{\varsigma}^{\tilde{\cal O}}(x)} \!(-\ip\delta_{\tilde{\cal O}}(\lambda^{\hat{\cal O}}{\bf F}_{\ {\cal O}{\cal O}'}^{ \hat{\cal O}}))\!+\!\!\ip{{\varsigma}^{\tilde{\cal O}}\!(x)}(-{\ip} C^{\ \hat{\cal O}}_{{\tilde{\cal O}},{{\cal O}''}}\lambda^{{\cal O}''}{\bf F}_{\ {\cal O}{\cal O}'}^{ {\cal O}''})\right)\right.\cdot\nn\\
&&\hspace{1cm}\cdot \left(\!\frac{2\ip}{4}\lambda_{\hat{\cal O}}(\ip{\bf F}^{\ {\cal O}{\cal O}'}_{\hat{\cal O}})\!\right)\left.-(\!-\ip\vartheta \varsigma^{\tilde{\cal O}}(x))(-\ip\delta_{\tilde{\cal O}}(L_o) \right)d\chi^n=0.\label{VarLoFS}
\eeaa
The term associated with the first term in the right-hand side of (\ref{VarLoFS}) cancels out with the term coming from the measure variation (\ref{VarMeasure}), both with the factor $\vartheta$—essentially the change of variables theorem. The term with the structure constant cancels due to the invariance of $L_o$ under global transformations. So, for arbitrary $\varsigma^{{\cal O}}(x)$—as long as the coordinate transformations remain bijective—we achieve invariance of the action $A_o$. Notice that we do not need the equations of motion for this invariance. Hence, under the given assumptions, both local gauge symmetry and transformations associated with the action of covariant derivatives are automatically satisfied. The picture that emerges is that, at least mathematically, the model originates from a Lie group manifold with its invariant Haar metric.

Later on, the task will be to extend this symmetry to a homogeneous group prior to a contraction delivering Poincar\'e-like translations. We contemplate in \cite{U1PoinExt} how exactly it evolves from an era involving such a homogeneous Lie-manifold dynamic to a subsequent era where some generators are "straightened" towards building a set of mutually commuting translations that dominate the directions in which covariant derivatives act.

\section{Alternative Considerations}
An alternative to the obtained non-contracted Lorentz extension model could be found if we entertain the notion where we want to direct the role of generators \(T_{\mu}, \bar{T}_{\nu}\) to be translations after a further process of contraction (after having performed the first round of contraction getting rid of the \(x^4\)). If we wanted to profile such a model as a forerunner of a gravity theory, the coupling constant \(\ell\) could take the form \(\lambda/ \Lambda\), where \(\Lambda\) is a dimension -1 parameter (a length) associated with a possible contraction, and \(\lambda\)  a dimensionless coupling constant.

In this case, we could view the fields \(\phi_a^{\ \mu}, \bar{\phi}_a^{\ \mu}\) as dimensionless vielbein fields, and the combination \({\bf g}_{ab}=\phi_a^{\,\mu}\bar{\phi}_{b\mu}\) will be a prospective complex metric (the symmetric part being real, the antisymmetric imaginary). But such a metric is only invariant in the contracted limit, which needs to be performed. Hence, a term analogous to \(\sqrt{-\text{det}({\bf g})}\) would be obtained from \(\frac{1}{4!}\epsilon_{\mu\nu\rho\sigma}\phi_{a}^{\ \mu}\hat{\phi}_{b}^{\ \nu}\phi_{c}^{\ \rho}\hat{\phi}_{s}^{\ \sigma}\epsilon^{abcs}\). Alternatively, in order to obtain a term like this from an actual global invariant prior to contraction, we appeal to the non-homogeneous hollow Casimir (\ref{casNLEHollow}). Using the similar transformation properties for field and field strengths, we could propose as a forerunner of the dimensionless \(\sqrt{-\text{det}({\bf g})}\) the Lagrangian density factor of the form:
\beaa
\epsilon_{\mu\nu\rho\sigma}\left({\frac{\lambda^4\epsilon^2}{4}}\phi_{a}^{\ \mu}\hat{\phi}_{b}^{\ \nu}\phi_{c}^{\ \rho}\hat{\phi}_{s}^{\ \sigma}+\Lambda^4 {r'^2}{{G}}^{[\mu\nu]}_{\ \ \ ab}{{G}}^{[\rho\sigma]}_{\ \ \ cs} \right)(1/4!)\epsilon^{abcs}. 
\eeaa
It is not clear, however, that this term can compensate for the variation of the measure \(dx^n\) under some of the transformations. We need to examine each symmetry transformation and the corresponding equation of motion separately. On the other hand, the degrees of freedom associated with fields \(\phi_a^{\ \mu}, \hat{\phi}_a^{\ \mu}\) need to be matched with a prospective complex scalar for the Higgs field and a complex metric. The insightful reader will notice that the quadratic term in the field strengths antisymmetric in their Lorentz (upper) indices resembles the chiral anomaly term, and might be instrumental in implementing a path--integral measure that might help address this obstruction. In \cite{U1PoinExt}, we include separately the forerunners of space--time translations prior to \(x^4\)--contraction. We will have a homogeneous hollow casimir (identity), which can involve a mixture of the field strength antisymmetric in their Lorentz (upper) indices or in indices associated with (will--be) internal (gauge) symmetries.

Consider now a model symmetric under the 10D algebra \(<\!\{M_{\mu\nu},T_{\rho}\!+\!\hat{T}_{\rho}\}\!>\). This will entail the restriction to the corresponding covariant derivative with: a real field \(\tilde{\phi}_a^{\ \rho}\) with coupling constant \(\lambda/\Lambda\) associated with \(T_{\rho}+\hat{T}_{\rho}\), and a real field \(B_a^{\,[\mu,\nu]}\) with coupling constant \(r'\) associated with \(M_{\mu\nu}\). The quadratic casimir (\ref{casdim10subalg}) leads to a globally invariant Lagrangian density along the same lines as done for the full LE. In this case, the generators involved in the hollow casimir (\ref{casdim10subalgHollow}) do include all symmetry generators. So, by taking as the prospective metric \({\bf g}_{ab}=\tilde{\phi}_a^{\,\mu}\tilde{\phi}_{b\mu}\), we could propose as a forerunner of the dimensionless \(\sqrt{-\text{det}({\bf g})}\) the Lagrangian density factor of the form:
\beaa
\epsilon_{\mu\nu\rho\sigma}\left({{\lambda^4\epsilon^2}}\tilde{\phi}_{a}^{\  \mu}\tilde{\phi}_{b}^{\ \nu}\tilde{\phi}_{c}^{\ \rho}\tilde{\phi}_{s}^{\ \sigma}+\Lambda^4 {r'^2}{{G}}^{[\mu\nu]}_{\ \ \  ab}{{G}}^{[\rho\sigma]}_{\ \ \ cs}\right)(1/4!)\epsilon^{abcs}. 
\eeaa
The globally invariant factors presented used \((1/4!)\epsilon^{abcs}\) to condensate the covariant derivative indices, but it might be omitted, and set \(a=0,\cdots, s=3\) that connect to the labels in the measure \(dx^n\). There are reasons for concern in using a non-homogeneous invariant identity.

Now, the contraction procedure used to arrive at the Poincar\'e algebra can also be considered for the variables involved in the space--time complexification. We could set \({\bf T}_{\mu\nu}=(1/\Lambda^2) {T}_{\mu\nu}, {\bf T}_{\rho}=(1/\Lambda)T_{\rho},\hat{\bf T}_{\rho}=(1/\Lambda)\hat{ T}_{\rho}\), which in the end corresponds to the replacement \(r=\tilde{r}/\Lambda^2, \ell=\lambda/\Lambda\). In fact, the contraction formalism, although a well-posed mathematical procedure for undergoing from a Lie algebra to another, does not necessarily account for the actual physical underpinnings for such a procedure. The nature of the gravitational coupling naively recalls the Fermi constant, which just accounts for the effective remnant of a massive gauge field. We do not know if this is the case here, but playing the game of introducing dimensionful couplings (and dimensionless fields) let us foresee possible scenarios for the path to effective low-energy symmetry.

The building of a factor embodying an invariant measure suggests a quartic invariant (perhaps extending the Pauli-Lubanski invariant) in the fields. Some extension of the quadratic invariant in (\ref{QuarticInv}) might lead to the ultimate invariant measure. 

\section{Relations, Outlook, Conclusions}
We have launched a program to profit from certain observations on structures of the SM of particle physics that aimed to obtain symmetries that render naturalness
to them. To simplify matters, we departed from a model  with just an internal $U(1)$ symmetry, and proposed a transformation combining Higgs and gauge fields inspired in their similar ultraviolet behavior, and their ultimate recombination during EW symmetry breaking.  In attempting to close an algebra for such a symmetry, we were led to a Lie algebra involving both internal $U(1)$ and external symmetries, such as Lorentz transformations. This extension renders by itself novel extensions of external symmetry involving a traceless spin 2 generator, some symmetric vectors, and scalars. For the obtained extension of the Lorentz algebra we constructed first a  locally invariant model  but without affecting the geometric frame (so, not actually involving external symmetry), where some gauge fields seem to adopt the role of a Higgs--like scalar field.  

 We  developed  representations of the symmetry in terms of differential operators that allowed to classify the extension as a pseudo--unitary Lie algebra of dimension 25, and gain insight into its geometrical counterpart. It also prepares the path for obtaining the corresponding Poincar\'e extensions in \cite{U1PoinExt}. We incorporate a differential representation into a model whose geometric frame is coordinatized by the variables involved in the differential representation. The differential operators reproduce locally the symmetry, and the local gauge transformations introduce  smooth distortions from such background. The labels of the covariant derivatives are those of the geometric manifold. 
 The considered extension do not have  operators to be  forcefully promoted to be the  ultimate space--time translations unless such a role is conferred to some of the generators in the extension.  Some preliminary considerations in alternative models  gave an   exploration of these possibilities. The application of an extended equivalence principle might grant the existence of a locally flat or de Sitter space--time measure. Our equations of motions were deprived from these constructions. We will tackle more extensively such considerations when dealing with Poincar\'e--like extensions, where there are clearly designated translations.

 These constructions reveal that the Lorentz  Lie algebra extension LE is the Lie algebra of ${\mathfrak{u}}(s_1,t_1)$ with  $s_1,t_1>0$ and $s_1+t_1=5$. In order to assign the origin of the Higgs field to certain vector generators, $t_1=1$ is demanded.  The obtained extension include as subalgebras some novel extensions of the Lorentz algebra  not considered in the literature elsewhere. In the differential representations we introduced a prospecting dimension (here $\chi^4$ and perhaps $y^4$) since this will enable dealing with Poincar\'e extensions in a further exploration in \cite{U1PoinExt}. As pointed out, the  Lorentz Extension (LE)   contains readily the seeds for gravity considerations. We  expect, that a linear combination of these translation precursors with the translations to be introduced in \cite{U1PoinExt} will compose the ultimate translations and its massless (or close to massless) graviton, while the further translation--like combination  breaks and their graviton--like field acquire significant mass. 

 Only a more accurate examination of the diverse possibilities, actual paths towards the low energy effective model of the Standard Model with gravity, and the hopefully, stability of involved symmetries under quantum fluctuations will help to settle which extensions might be more adequate. It is not clear from the outset, which will be the symmetry breaking path or contraction procedures leading at low energies to the aimed features of the SM.  A following issue of this program in \cite{U1PoinExt} will obtain the corresponding extension of the Poincar\'e symmetry and  address the introduction of gravity through the local invariance under forerunners of local space--time translations. In fact, the transit from the models with local symmetry prior contraction to models after contraction seems to model inflation  since then certain generators become mutually commuting translations which overwhelms all other covariant derivatives (suppressed by factors $1/R$, for $R$ the contraction parameter.)  The LE  algebra obtained here will be used for an extended spin connection in an extended General Relativity theory with corresponding contracted Poincar\'e Extension. Now, how do we enlarge such simplified models to include isospin ($SU(2)$) and strong interaction ($SU(3)$) internal symmetries will be  tackled in further contributions, first aiming to include the  SM electroweak bosonic sector, and then tackling larger internal symmetries. 

The complexification underlying the so called 10D differential representation seems to converge in part to several developments leading to similar features. From the early unification attempts by Einstein and Strauss \cite{EinsteinStrauss}, and E. Schr\"odinger \cite{ESchr}, to  diverse remarkable constructions in the crossroads of string theory, non--commutative geometry,  and gravity theory  \cite{TorresdC}\cite{Hughston}\cite{Witten}\cite{Ashtekar}\cite{PST}\cite{Chamseddine2001}\cite{GORS}
\cite{Chamseddine2004}\cite{Valtancoli}\cite{Esposito}\cite{Bogos}\cite{Chamseddine2006}\cite{LiSotiriouBarrow}\cite{SotiriouLieBarrow}\cite{HeckmanVerline}\cite{ChamConnes}\cite{MMZ}. Some  have implemented complexifications to cope with diverse challenges at Planck scales, compactification, tenets of noncommutative geometry, and phenomenological results on Higgs mass. In particular, the 15 dimensional subalgebra $l=<\{M_{\mu\nu},T_{\rho\sigma}\}>$ in (\ref{MinExStart})--(\ref{MinExEnd}) could coincide with the algebra ${\mathfrak{so}}(1,5)$ (or the conformal ${\mathfrak{so}}(2,4)$) used in Yang's Space--time and its modified version \cite{Yang}\cite{Tanaka}, and the one used to construct a fuzzy  4D de Sitter space--time in \cite{HeckmanVerline}\cite{MMZ}. The non contracted  Lorentz extension provide a novel way to produce a fuzzy 4D de Sitter space--time. The geometrical frame underlying the differential representations and model will profit from  prominent constructions dealing with  compactification of extra (now hidden) dimensions \cite{CMJR}\cite{RS1}\cite{RS2}. The present approach  seems to extend noncommutative geometry constructions where the Higgs fields is the gauge field associated to discrete dimensions (or spectra) and seemingly discrete algebraic structures. Here, the would--be--Higgs field  proceeds  from a gauge field of a continuous symmetry. It might be the case that symmetry breaking, contractions or emerging brane structures constrain the symmetry associated to such Higgs field to converge with the tenets of noncommutative geometry  models. It is appealing, that similar or identical substructures have emerged in so diverse settings. Some constructions and evolutions that are natural in some settings might be intricate and almost ad--hoc in other approaches. That is why novel model building settings are welcome, even if some features have been revealed elsewhere.

One remark on noncommutative and/or nonassociative coordinates is in place: The use of noncommutative  and even nonassociative coordinates as parameters of transformations that composes ( via parameter times generator)  Lie group structures can be avoided in some instances. When the noncommutativity and/or nonassociativity  is governed by (discrete) finite gradings compatible with the Lie algebraic structure, and there exist a finite basis embodying in the parameters such noncommutative and nonassociative behaviour, then there is a change of variables (parameters) restoring commutativity and associativity \cite{GNCNA}. Under these hypothesis, this  leaves  only  Grassmann variables noncommutativity (and the corresponding super Lie algebras structures) as  the only truly nontrivial contributions in the noncommutative and nonassociative realm. If the referred  change of variables can be performed locally in the space--time, and there is a smooth  change of variables for neighboring space--time points, then the use of noncommutativity and/or nonassociativity might be avoided if a smooth coordinate transformation could be pasted  for the whole space--time. The existence of transformations maintaining self--duality needs to be settled separately. Nevertheless, there is to my knowledge, no general assessment of the possibility of having such variable changes. There is in Noncommutative Geometry an unexpected richness of structures rendering some  naturality to  the origin of multiple features of the SM and gravity. 

The phenomenological corrections to Higgs decays into gauge bosons have been extensively studied in various Beyond Standard Models (BSM) \cite{Aadetal}. These models encompass scenarios where the Higgs is considered as a composite Higgs boson \cite{CompH}, a pseudo Nambu--Goldstone boson \cite{Caoetal}, a component of the Next--to--Minimal Supersymmetric Standard Model (NMSSM) \cite{NMSSM},  in the Gauge--Higgs Unification scheme (GHU) \cite{GHU}, or even as a neutral scalar Higgs imposter \cite{Lowetal}. The experimental deviation from the Standard Model (SM) branching ratio for the decay of the Higgs to a Z boson and a photon, as reported in \cite{Aadetal}, presents a promising avenue for testing and constraining BSM scenarios. The models developed in this work, in \cite{U1PoinExt}, and in forthcoming contributions with an internal symmetry of $U(1)\times SU(2)$ might contribute to the exploration of scenarios predicting enhanced branching ratios for Higgs decays into gauge bosons.

We will pursue an extension of symmetry  involving fermionic generators, so that we can aim to assign fermionic particles to remnants of such generators, joining early visionary developments by P. Dirac  \cite{Dirac}. Some developments in quantum gravity on entropy have assigned a quantum of area in terms of squared Planck length to a black hole  event horizon \cite{Beke}\cite{BH} or a further inner horizon singularity \cite{CNC}. It might be that it is the exclusion principle in place assigning some fundamental fermionic particles to diverse quantum area locations in the ultimate horizon.
 
\section{Acknowledgments} The first stages of this research took place at the Departamento de F\'{\i}sica Te\'orica y del Cosmos, Universidad de Granada, Spain, and CIMAT, Guanajuato, M\'exico. The Fundaci\'on Carolina is acknowledged for the financial support of the research stay in Spain, and CIMAT is acknowledged for the finantial support for the research stay in M\'exico. The Universidad Nacional de Colombia, Sede Medell\'{\i}n, is aknowleged for its support of this research.

\end{document}